\documentclass[preprint, 11pt, authoryear, a4paper]{elsarticle}
\usepackage[margin=2.5cm]{geometry}
\usepackage{setspace}

\usepackage{amssymb}
\usepackage{amsmath}
\usepackage{amsthm}
\usepackage{stmaryrd}
\usepackage{booktabs}
\usepackage{multirow}
\usepackage{color}
\usepackage[dvipsnames]{xcolor}
\usepackage[table,xcdraw]{xcolor}
\usepackage{svg}
\usepackage{todonotes}
\usepackage{siunitx}
\usepackage{tikz}
\usepackage{pgfplots}
\usepackage{graphicx}
\usepackage{lscape}
\usepackage[linesnumbered,ruled,vlined]{algorithm2e}
\usepackage[breaklinks]{hyperref}
\SetKwInput{KwInput}{Input}                
\SetKwInput{KwOutput}{Output}              
\usetikzlibrary{
    decorations.pathreplacing, 
    angles, 
    quotes, 
    matrix, 
    fit, 
    positioning, 
    arrows.meta,
    decorations.pathmorphing, 
    decorations.shapes,
    shadings,
    fadings,
}
\usepackage{pifont}

\newcommand\singleIntervals{\mathbb{I}_1}
\newcommand\singleInterval{v}
\newcommand\intervals{\mathbb{I}}
\newcommand\multipleInterval{w}
\newcommand\intervalSingles{\mathcal{S}}
\newcommand\intervalShape{\mathcal{D}}
\newcommand\shapes{D}
\newcommand\shape{d}

\newcommand\nodes{N}
\newcommand\nodeIdx{n}
\newcommand\receivers{R}
\newcommand\receiverIdx{r}
\newcommand\plan{\mathit{T\!P}}
\newcommand\plans{P}
\newcommand\planIdx{p}
\newcommand\timesteps{Q}
\newcommand\timestepIdx{q}

\newcommand\tracks{T}
\newcommand\track{t}
\newcommand\trackIdx{i}
\newcommand\emitters{E}
\newcommand\emitter{e}
\newcommand\emitterIdx{k}
\newcommand\maxBandwidth{b}
\newcommand\surveys{S}
\newcommand\survey{s}
\newcommand\surveyIdx{j}

\newcommand\object{o}
\newcommand\task{x}

\newcommand\subsurveys{U}
\newcommand\subsurvey{u}
\newcommand\subsurveyIdx{l}

\newcommand\freq{\omega}
\newcommand\freqLow{f'}
\newcommand\freqUp{f''}
\newcommand\frequency{f}

\newcommand\goalObservation{\mathit{G\!R}}
\newcommand\realizedObservation{\mathit{R\!R}}
\newcommand\currentObservation{\mathit{C\!R}}

\newcommand\obj{\Theta}

\newcommand\splitSize{\psi}

\newcommand\configurations{C}
\newcommand\configuration{c}
\newcommand\goalInsertion{\mathit{I\!R}}
\newcommand\weight{w}

\newcommand\complexity{\zeta}

\newcommand\parentCount{\rho}
\newcommand\variableCount{\alpha}
\newcommand\constrainCount{\beta}
\newcommand\precision{\Delta}

\newcommand\queue{L}
\newcommand\positions{G}
\newcommand\position{g}

\newcommand\historicalObservation{\mathit{H\!R}}
\newcommand\discount{\gamma}

\newcommand\utilization{\mu}
\newcommand\utilizationTrack{\utilization^T}
\newcommand\utilizationSurvey{\utilization^S}
\newcommand\proportion{\lambda}

\newcommand\balance{\xi}
\newcommand\priority{\Xi}

\journal{Computers and Operations Research}

\begin{document}

\begin{frontmatter}



\title{Fast Resource Management Algorithm for Passive Surveillance Systems} 

\author[aff:ciirc,aff:fee]{Jan Pikman\corref{ca}}
\ead{pikmajan@fel.cvut.cz}
\author[aff:ciirc]{Přemysl Šůcha}
\ead{premysl.sucha@cvut.cz}
\author[aff:vut,aff:era]{Jerguš Suja}
\ead{j.suja@era.aero}
\author[aff:era]{Pavel Kulmon}
\ead{p.kulmon@era.aero}
\author[aff:ciirc]{Zdeněk Hanzálek}
\ead{zdenek.hanzalek@cvut.cz}
\affiliation[aff:ciirc]{
    organization={Czech Institute of Informatics, Robotics and Cybernetics, Czech Technical University in Prague},
    country={Czech Republic}
}
\affiliation[aff:fee]{
    organization={Faculty of Electrical Engineering, Czech Technical University in Prague},
    country={Czech Republic}
}
\affiliation[aff:vut]{
    organization={Faculty of Mechanical Engineering, Brno University of Technology},
    country={Czech Republic}
}
\affiliation[aff:era]{
    organization={Research Department, ERA a.s.},
    city={Pardubice},
    country={Czech Republic}
}
\cortext[ca]{Corresponding author at: Czech Institute of Informatics, Robotics, and Cybernetics, Czech Technical University in Prague, Jugoslávských partyzánů 1580/3, 160 00 Prague 6, Czech Republic.}

\begin{abstract}
    Passive surveillance systems (PSS) detect and track objects that emit electromagnetic signals from hundreds of kilometers away.
    These systems have a limited number of receivers and can only observe a fraction of the frequencies of interest simultaneously.
    To improve its behavior, we propose the ResourceTune algorithm, which iteratively constructs optimized schedules to determine which frequencies each receiver should observe at a given time step.
    The algorithm's main component is the optimization of receiver configurations using a left-right heuristic combined with linear programming.
    Our approach is unique because, unlike others, we focus on optimizing available resources and observed frequencies, which was never done before.
    We experimentally compared the proposed algorithm with a greedy and the state-of-the-art method for construction of PSS schedules. 
    In most of the considered scenarios, ResourceTune outperformed both algorithms, and in the most extreme case, its objective value was more than $2.7$ times better than the values reached by other methods.
\end{abstract}


\begin{highlights}
    \item Focus on an unexplored area of resource management of passive surveillance systems
    \item Resource optimization is achieved by novel optimization of observed frequency bands
    \item The proposed algorithm is fast, making it applicable in the real world
    \item Significantly better results than the current state-of-the-art method
\end{highlights}


\end{frontmatter}


\newpage
\section{Introduction}

\begin{figure}
    \centering
    \begin{tikzpicture}
        \node[anchor=south west,inner sep=0] at (0,0) {
            \includegraphics[page=2, clip, trim=9cm 1cm 0.5cm 21.75cm, width=0.75\textwidth]{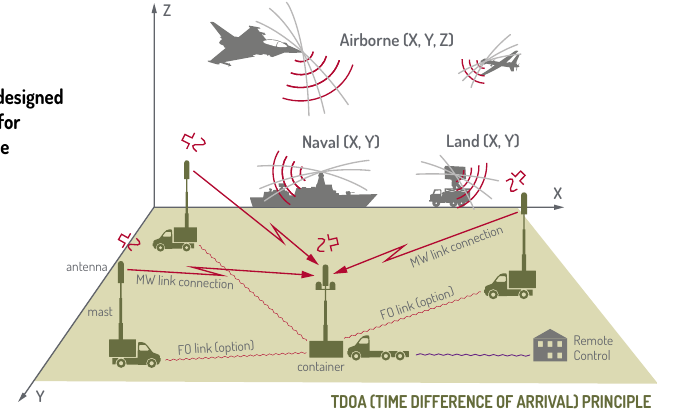}
        };
        \fill[fill=white] (-0.05,4) rectangle (2,7);
    \end{tikzpicture}
    \caption{
        Illustration of the working principles of PSS. 
        Image taken from \citep{era}.
    }
    \label{fig:era}
\end{figure}

A passive surveillance system (PSS) can be described as a complex electronic support measure that intercepts broad-spectrum electromagnetic pulses emitted by various sources in the air, on land, and at sea, collectively called targets.
Based on these interceptions the PSS is able to detect, localize, track, and possibly even identify these targets.
This is illustrated in \figurename~\ref{fig:era}.
As can be seen, its functionalities are practically identical to those of traditional radar systems, with the advantage that its operation is covert since, unlike radars, it does not need to emit electromagnetic pulses into the environment.
Therefore, it is easier for the PSS to remain hidden from potential observers.

For the PSS to function properly and perform all of the expected tasks, which include surveillance, detection, identification, and tracking of targets, it is essential to  manage its limited resources correctly.
The resources are receivers that observe the frequencies and thus perform the tasks.
These observed frequencies are determined by configurations of the receivers. 
Therefore, it is important to select the suitable receiver configurations at each point in time to ensure that the PSS functions properly and efficiently.
In addition, the PSS should be able to swiftly adapt its behavior because of the ever-changing environment where targets can quickly appear and disappear.
As a result, the PSS resource management algorithm must operate in real time.

This paper presents a unique perspective on a currently almost unexplored field of resource management of PSSs (PSSRM), focusing on the optimization of available resources and observed frequencies.
We consider a PSS that realizes two types of tasks, track and survey, determining which frequency bands to observe and how often.
The goal is to realize as many of these tasks as possible, as well as possible.
To solve this problem, we propose an algorithm, called ResourceTune, that decomposes the problem into multiple steps that can be solved in real time.
It uses a left-right heuristic to construct receiver configurations that determine which frequency bands the receiver observes, so that each configuration realizes as many tasks as possible.
Then, we use linear programming (LP) to determine how often the receivers should be set to a particular configuration.
We compare ResourceTune in several different scenarios with the greedy algorithm, which is inspired by the typical radar resource management (RRM) algorithm called time balancing algorithm \citep{stafford_mesar, butler_time_balancing}, and the PSSRM state-of-the-art algorithm proposed by \citet{suja_scheduling}.

\subsection{Contributions}

The main contributions of this paper can be summarized as follows:
\begin{enumerate}
    \item 
        \emph{The paper focuses on the almost unexplored topic of PSSRM.}
    \item 
        \emph{The proposed algorithm has low computational complexity} and is therefore suitable for direct use in real-world applications.
    \item 
        \emph{The introduced left-right heuristic provides an elegant compromise between computational complexity and optimality.} It greatly simplifies the selection of receiver settings, which would otherwise be almost unmanageable.
    \item 
        \emph{To the best of our knowledge, the frequency optimization of receiver configurations has never been done before} and could be applied to other research areas that are concerned with the optimization of frequency coverage.
    \item 
        As mentioned above, \emph{we demonstrate the superiority of ResourceTune by comparing its behavior with the PSSRM state-of-the-art algorithm introduced by \citet{suja_scheduling}}.
\end{enumerate}
Furthermore, our algorithm was developed in close cooperation with the industry and with a specific system in mind, which underscores the applicability of our approach.

\subsection{Paper Outline}

The rest of this paper proceeds as follows.
Section~\ref{sec:preliminaries} describes preliminaries, most of which are connected to multiple-intervals.
In Section~\ref{sec:problem_statement}, a problem description is given, including a simplified description of the considered PSS.
An overview of the related literature is provided in Section~\ref{sec:related_work}.
The ResourceTune algorithm is introduced and described in detail in Section~\ref{sec:methodology}.
Section~\ref{sec:experiments} describes the experimental results.
Section~\ref{sec:conclusion} summarizes the paper and discusses the presented results.


\section{Preliminaries}
\label{sec:preliminaries}

Throughout this paper, we often mention frequency bands and their unions.
To simplify their description, we w.l.o.g. omit the units and introduce the following interval notation.
The interval defined as $ \singleInterval = [\freqLow, \freqUp] = \{\frequency \in \mathbb{R}_{>0} \;|\; \freqLow \leq \frequency \leq \freqUp\}$, where $\freqLow, \freqUp \in \mathbb{R}_{>0}$ and $\freqLow < \freqUp$, will be called \emph{single-interval}.
The size of the single-interval $|\singleInterval|$ is equal to $\freqUp - \freqLow$.
The set of all single-intervals is denoted by $\singleIntervals$.

The union of an arbitrary number of pairwise disjoint single-intervals is called a \emph{multiple-interval}.
The set of all multiple-intervals is denoted by $\intervals$.
Consider a multiple-interval $\multipleInterval$; the set of single-intervals it consists of is denoted by $\intervalSingles(\multipleInterval)$.
The size of multiple-interval $|\multipleInterval|$ is equal to $\sum_{\singleInterval \in \intervalSingles(\multipleInterval)} |\singleInterval|$.
The \emph{shape} of a multiple-interval $\intervalShape(\multipleInterval)$ is an odd-length vector of positive real numbers that describes the sizes of its constituent single-intervals and the spaces between them, from left to right. 
For example, consider multiple-interval $\multipleInterval =[0, 5] \cup [7,8] \cup [11, 15]$, then $\intervalSingles(\multipleInterval) = \{[0, 5], [7,8], [11, 15]\}$ and $\intervalShape(\multipleInterval) = (5, 2, 1, 3, 4)$.
We define the set of all multiple-intervals induced by set of shapes $\shapes$ as $\intervals[\shapes] := \{\multipleInterval \in \intervals \,|\, \intervalShape(\multipleInterval) \in \shapes\}$.


\section{Problem Statement}
\label{sec:problem_statement}

\begin{figure}
    \centering
    \includegraphics[width=\textwidth]{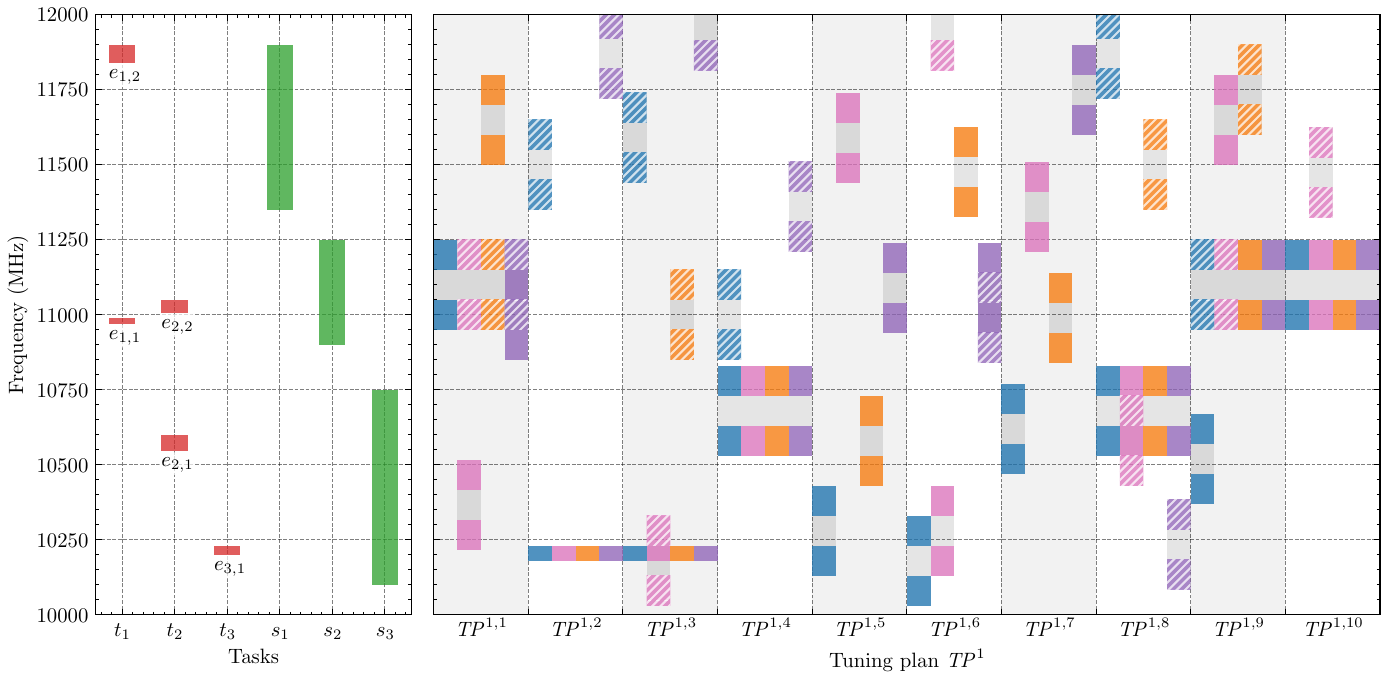}
    \caption{
        The left part of the figure shows the frequency bands of the individual tasks. 
        Their complete properties are described in Table~\ref{tab:tasks}.
        The right side of the figure shows tuning plan $\plan^1$.
    }
    \label{fig:problem_statement}
\end{figure}

We assume the following PSS, which is an obfuscated version of an existing PSS that is actively in use.
It consists of four identical sensor nodes $\nodes = \{1, 2, 3, 4\}$ indexed by $\nodeIdx$.
These can be seen in \figurename~\ref{fig:era}.
Each sensor node contains two identical receivers $\receivers = \{1, 2\}$ indexed by $\receiverIdx$.
The frequency bands observed by the receivers over time are determined by consecutive tuning plans $\plan^1, \dots, \plan^{|\plans|}$. 
The tuning plans are indexed by $\planIdx \in \plans = \{1, \dots, |\plans|\}$.
Each tuning plan $\plan^{\planIdx}$ specifies the behavior of each receiver for consecutive time steps indexed by $\timestepIdx \in \timesteps = \{1, \dots, |\timesteps|\}$. 
Receiver $\receiverIdx$ on sensor node $\nodeIdx$ during time step $\timestepIdx$ of tuning plan $\planIdx$ observes frequency bands corresponding to multiple-interval $\plan_{\nodeIdx, \receiverIdx}^{\planIdx, \timestepIdx} \in \intervals[\shapes]$, where $\shapes$ is a set of allowed receiver shapes, i.e., the multiple-intervals the receiver can observe simultaneously.
An example of a tuning plan is shown in \figurename~\ref{fig:problem_statement}.

There are two types of tasks: track and survey.
We consider \emph{tracks} $\tracks = \{\track_{1}, \dots, \track_{|\tracks|}\}$, each corresponding to one object tracked by the PSS.
Each track $\track_{\trackIdx}$ has emitters $\emitters_{\trackIdx} = \{\emitter_{\trackIdx, 1}, \dots, \emitter_{\trackIdx, |\emitters_{\trackIdx}|}\}$.
Emitter $\emitter_{\trackIdx, \emitterIdx}$ has a frequency band represented by single-interval $\freq_{\emitter_{\trackIdx, \emitterIdx}}$, where the emitter broadcasts, and maximum bandwidth $\maxBandwidth_{\emitter_{\trackIdx, \emitterIdx}} \in \mathbb{R}_{\geq 0}$, which limits the receiver's bandwidth when observing the emitter.
This reduces the amount of noise measured by the receiver.
Emitter $\emitter_{\trackIdx, \emitterIdx}$ is \emph{observed} at time step $\timestepIdx$ of tuning plan $\planIdx$ if, at that time, there is at least one receiver on each sensor node that observes the same multiple-interval $\multipleInterval \in \intervals[\shapes]$ and emitter's frequency band $\freq_{\emitter_{\trackIdx, \emitterIdx}}$ is a subset of single-interval $\singleInterval \in \intervalSingles(\multipleInterval)$ while $|\singleInterval| \leq \maxBandwidth_{\emitter_{\trackIdx, \emitterIdx}}$.
Track $\track_{\trackIdx}$ also has goal observation rate $\goalObservation_{\track_{\trackIdx}} \in [0, 1]$ that specifies the proportion of time steps when the target should be observed.
The track is \emph{observed} at time step $\timestepIdx$ of tuning plan $\planIdx$ if one of its emitters is observed at that time.

\emph{Surveys} $\surveys = \{\survey_1, \dots, \survey_{|\surveys|}\}$ specify at which frequencies to search for new objects.
Survey $\survey_\surveyIdx$ is characterized by a frequency band of interest represented by single-interval $\freq_{\survey_{\surveyIdx}}$ and goal observation rate $\goalObservation_{\survey_\surveyIdx} \in [0, 1]$ that specifies the proportion of time steps each of the frequencies belonging to $\freq_{\survey_\surveyIdx}$ should be observed.
Frequency $\frequency \in \freq_{\survey_\surveyIdx}$ is \emph{observed} at time step $\timestepIdx$ of tuning plan $\planIdx$ if, at that time, at least one receiver on any sensor node observes frequency $\frequency$.
Both track and survey frequency can be observed \emph{only once per time step}.
An example of various tasks is shown in \figurename~\ref{fig:problem_statement}.

\begin{table}[]
\centering
\caption{Overview of tasks shown in \figurename~\ref{fig:problem_statement}.}
\label{tab:tasks}
\begin{tabular}{@{}rrrrr@{}}
\toprule
\multicolumn{1}{c}{task} & \multicolumn{1}{c}{$\goalObservation_\cdot$} & \multicolumn{1}{c}{$\emitter_{\cdot, \cdot}$} & \multicolumn{1}{c}{$\freq_{\cdot}$} & \multicolumn{1}{c}{$\maxBandwidth_{\emitter_{\cdot, \cdot}}$} \\ \midrule
\rowcolor[HTML]{EFEFEF} 
\cellcolor[HTML]{EFEFEF} & \cellcolor[HTML]{EFEFEF} & $\emitter_{1,1}$ & $[10970, 10990]$ & 100 \\
\rowcolor[HTML]{EFEFEF} 
\multirow{-2}{*}{\cellcolor[HTML]{EFEFEF}$\track_1$} & \multirow{-2}{*}{\cellcolor[HTML]{EFEFEF}0.3} & $\emitter_{1,2}$ & $[11840, 11900]$ & 100 \\
 &  & $\emitter_{2,1}$ & $[10545, 10600]$ & 100 \\
\multirow{-2}{*}{$\track_2$} & \multirow{-2}{*}{0.5} & $\emitter_{2,2}$ & $[11005, 11050]$ & 100 \\
\rowcolor[HTML]{EFEFEF} 
$\track_3$ & 0.2 & $\emitter_{3,1}$ & $[10200, 10230]$ & 50 \\
$\survey_1$ & 0.4 & --- & $[11350, 11900]$ & --- \\
\rowcolor[HTML]{EFEFEF} 
$\survey_2$ & 0.5 & --- & $[10900, 11250]$ & --- \\
$\survey_3$ & 0.3 & --- & $[10100, 10750]$ & --- \\ \bottomrule
\end{tabular}
\end{table}

Our goal is to construct tuning plans $\plan^1, \dots, \plan^{|\plans|}$ so that tasks are observed as often as stated by their goal observation rate.
Since we consider the observation of tracks and surveys equally important, we define the objective function as follows:
\begin{align}
    \label{eq:obj}
    & \min_{\plan^1, \dots, \plan^{|\plans|}} \obj = \min_{\plan^1, \dots, \plan^{|\plans|}} \sum_{\track_\trackIdx \in \tracks} \obj_{\track_\trackIdx} + \sum_{\survey_\surveyIdx \in \surveys} \obj_{\survey_\surveyIdx},
\end{align}
where $\obj_{\track_\trackIdx}$ and $\obj_{\survey_\surveyIdx}$ represent how well track $\track_\trackIdx$ and survey $\survey_\surveyIdx$ are observed, respectively.
These values are defined:
\begin{align}
    \obj_{\track_\trackIdx} &= \max \left\{ 0, \goalObservation_{\track_\trackIdx} - \frac{1}{|\plans|} \sum_{\planIdx \in \plans} \realizedObservation_{\track_\trackIdx}^\planIdx \right\},
    \\
    \obj_{\survey_\surveyIdx} &= \frac{1}{|\freq_{\survey_\surveyIdx}|} \int_{\freq_{\survey_\surveyIdx}} \max \left\{ 0, \goalObservation_{\survey_\surveyIdx} - \frac{1}{|\plans|} \sum_{\planIdx \in \plans} \realizedObservation_{\frequency}^\planIdx \right\} d\frequency,
\end{align}
where $\realizedObservation_{o}^\planIdx = \frac{1}{|\timesteps|} \sum_{\timestepIdx \in \timesteps} \llbracket \object \text{ is observed by } \plan^{\planIdx, \timestepIdx} \rrbracket$ denotes the proportion of time steps in which observable object~$\object$, e.g., track $\track_\trackIdx$ or frequency $\frequency$, is observed during tuning plan $\planIdx$.

This problem statement assumes that the tasks will remain the same for $|\plans|$ consecutive tuning plans. 
In practice, however, tasks may change due to a rapidly changing environment in which new objects of interest can emerge quickly.
Therefore, we assume that the number of consecutive tuning plans $|\plans|$ for which the tasks do not change is unknown.
This means that tuning plans cannot be built all at once, but rather must be constructed one by one.
Consequently, each tuning plan must be constructed during the execution of the previous one, which creates a non-trivial time limit.

Figure~\ref{fig:problem_statement} illustrates our problem.
The left side of the figure shows examples of tasks.
There are three tracks $\tracks = \{\track_1, \track_2, \track_3\}$.
The frequency bands corresponding to their emitters are shown in red.
The figure also shows three surveys $\surveys = \{\survey_1, \survey_2, \survey_3\}$ whose frequency bands are colored green.
Table~\ref{tab:tasks} lists all of the tasks' properties in detail.
The right part of the figure shows tuning plan $\plan^1$, which consists of $10$ consecutive time steps.
Furthermore, in this example the allowed receiver shapes are $\shapes = \{(y) \,|\, y \in \mathbb{N}: 10 \leq y \leq 100\} \cup \{(100, 100, 100)\}$.
The colored areas in the tuning plan correspond to the frequency bands observed during each time step.
There are four different colors representing four different sensor nodes, and the two patterns allow us to differentiate between receiver number $1$ and $2$.

As can be seen, emitter $\emitter_{1,1}$ is observed during the first, ninth, and tenth time steps, while emitter $\emitter_{1, 2}$ is never observed.
Consequently, track $\track_1$ is observed three times, which is equivalent to its goal observation rate $\goalObservation_{\track_1} = 0.3$.
Emitter $\emitter_{2,1}$ is observed during the fourth and eighth steps.
Emitter $\emitter_{2,2}$ is observed during the same time steps as $\emitter_{1,1}$, so it is observed three times.
Note that multiple emitters can be observed at once.
Overall, track $\track_2$ is observed five times.
Since emitter $\emitter_{3,1}$ has $\maxBandwidth_{\emitter_{3,1}}$ equal to $50$, receivers with a configuration that has shape $(100, 100, 100)$ cannot observe it.
For this reason, it is observed by receivers whose configuration is $(50)$ during the second and third time step.
Consequently, track $\track_3$ is observed twice.
Now, we will focus on the surveyed frequencies.
It can be seen that frequency $11750~\si{MHz}$ is observed four times, i.e., during time steps $1$, $2$, $8$, and $9$.
Similarly, frequency $11000~\si{MHz}$ is observed during time steps $1$, $5$, $6$, $9$, and $10$.
Upon further inspection, it is noticeable that the presented tuning plan fully observes all tasks, even through not all receivers are active during certain time steps, e.g., steps number $1$, $2$, and $7$.
This means that the objective value of this tuning plan is~$0$.



\section{Related Work}
\label{sec:related_work}

To the best of our knowledge, there is very little scientific literature on the design, improvement, or optimization of PSSRM.
This is likely because of the significant complexity of PSSs and their military applications.
Three articles focusing on the resource management of the VERA-NG PSS were recently published \citep{kulmon_suja_sensor, suja_scalarization, suja_scheduling}.
The articles address the task of jointly optimizing the target search and tracking by determining which frequency bands should be observed by the available receivers and when.
\citet{kulmon_suja_sensor} formulate this by a complex multi-criteria objective function based on the expected information gain of the targets and the optimal surveillance distribution.
This challenging problem is solved using the Non-dominated Sorting Genetic Algorithm II (NSGA-II) \citep{nsga-ii}.
The authors also propose solving the problem using the $\epsilon$-constrained method and genetic algorithm (GA).
\citet{suja_scalarization} transform the problem using goal programming scalarization and then solve it using GA.
In the most recent article \citep{suja_scheduling}, the authors reformulated the objective functions and their interactions, producing a problem formulation without parameters.
\citet{jiang_passive_location} optimized the behavior of a PSS to improve its target tracking.
They formulated the problem using an objective function that balances tracking accuracy, priority, and resource utilization.
They propose to solve this problem using GA with specialized operators.

The RRM is the research area that is currently the closest to the PSSRM.
Their similarity stems from their common goals of surveillance, detection, identification, and tracking of targets.
Unfortunately, the similarities mostly end there, as the systems work on very different principles, e.g. radars localize a target based on the time difference between signal transmission and reception, while PSSs use multilateration.
Therefore, the algorithms presented below are mostly inspirational, and their adaptation for use in PSSs is either completely pointless or extremely difficult.

In recent decades, RRM has become a prominent part of radar research~\citep{charlish_rrm}.
The term encompasses the automatic management of various radar tasks and the setting of almost all of possible radar parameters that affect its behavior to meet its operational requirements.
Many of these task management and parameter setting problems have different computational complexity and update periods.
For example, target identification takes a long time but only needs to be done once, while the decision of which target to scan next is repeated hundreds of times per second.
It is natural to execute these problems hierarchically based on their run time and update period.
Using these criteria, \citet{charlish_rrm} divide the field of RRM into three core components: (i) priority assignment, (ii) task management, and (iii) scheduling.

(i) \emph{Priority assignment}, sometimes called task prioritization, is the assignment of priority values or levels to individual tasks. 
These tasks may represent requests for area surveillance, initiation of new targets, updates of existing tracks, and more.
A task's priority often determines how many resources it should get, whether it should be selected more often during scheduling, or how important its tracking accuracy is~\citep{hashmi_ai_survey}.
Some of the RRM algorithms use priorities only as weights that influence the objective function of optimized problems \citep{qu_est_mc, shaghaghi_b&b, gaafar_rl_mcts}, while some use them in a hierarchical way \citep{orman_scheduling_coupled_jobs, moo_scheduling_two_slope}. 
Most existing papers agree that priority assignment algorithms should be based on expert knowledge with clear reasoning \citep{charlish_rrm, miranda_knowledge_rm, sherwani}, and that an operator must be able to easily override assigned priorities.
In their survey of RRM, \citet{ding_survey_rrm} mentions that at least two papers have proposed assigning target priorities using simple neural networks whose input is based on the known information about the target, such as its speed, direction, acceleration, and distance from the radar.
Similar information was used by \citet{miranda_knowledge_rm} who used a decision tree with fuzzy rules and variables to determine the priority of the target.

(ii) \emph{Task management} is the second component of RRM. 
It is responsible for selecting radar parameters and assigning resources to tasks according to their priority \citep{charlish_rrm}.
Therefore, it is sometimes referred to as parameter selection or resource allocation.
Task management is often formulated as an optimization problem to maximize the accuracy of tracked targets under the constraints imposed by the limited resources of the considered radar system.
Because it is run less frequently, these problems can take longer to optimize and can be more complex.
Many papers focus on constructing the best possible tracking objective function that models the future expected track accuracy 
\citep{zhang_resource_parn, shi_transmit, shi_multidomain}.
This approach is prominent in works focused on resource allocation and parameter selection in multi-radar systems (MRS), where the goal is to coordinate multiple radars at different locations to perform their tasks better than if they worked individually.

In contrast, \citet{vaillaud_simple} focus on detecting a single target using a single radar.
They formulate a complex objective function that maximizes the probability of detecting a moving target and optimize it using an iterative algorithm based on Brown's recursion \citep{brown_iteration}, which uses the partial solutions obtained by a greedy algorithm proposed in \citep{stone_optimal}.
The follow-up paper \citep{vaillaud_overlap} considers the same problem, except that the observed spaces can overlap.
They solve this using the Forward And Backward algorithm coupled with a subproblem solver based on either dynamic programming or a greedy approximation heuristic.

Another resource allocation concept with a significant presence in the literature \citep{ing_thesis, irci_qram_solutions, charlish_auction} is the so-called Quality of Service based Resource Allocation Model (Q-RAM).
It approximates the usually complex tracking performance objective function with an exponential one, resulting in a notable simplification of the model \citep{ing_thesis}.

(iii) \emph{Scheduling} is the final core component of RRM that determines the behavior of the radar by planning which tasks should be performed at what time, often according to their priority, duration, or time preference \citep{hashmi_ai_survey}.
As mentioned earlier, scheduling algorithms must be fast because the task durations are in the tens of milliseconds and pre-planning is not possible due to the rapidly changing operational situation.
Therefore, radar scheduling mostly uses trivial algorithms such as list scheduling with priority queues based on hand-crafted heuristics, earliest deadline first scheduling, or earliest start time (EST) algorithm \citep{hashmi_ai_survey, qu_est_mc, sherwani, orman_scheduling_coupled_jobs, butler_time_balancing}.

Shaghaghi and Adve \citep{shaghaghi_b&b} study scheduling with tasks that can be delayed at the cost of increasing the objective.
They solve this problem optimally with an adapted branch and bound (B\&B) algorithm.
Unfortunately, the B\&B algorithm is notoriously slow and thus impractical for direct use in RRM.
This problem is addressed in a follow-up paper \citep{shaghaghi_value_network}, where the algorithm is accelerated by pruning the search space according to the estimates produced by a neural network.
Next, \citet{shaghaghi_mcts} consider a similar problem but with more complex constraints and solve it using the Monte Carlo tree search (MCTS) method. 
\citet{gaafar_rl_mcts} further improve this approach by modifying MCTS to ignore fully explored branches and by using reinforcement learning methods to improve the learning of the policy network.


Almost all of the algorithms presented in this section have properties that make them unsuitable for our problem.
Many of these algorithms focus solely on target tracking, ignoring the search for new targets.
The scheduling algorithms assume that tasks have specific times at which they should be scheduled.
For PSS, this would mean that the system knows precisely when the target emits. 
This is an extremely complex task, given the limited literature on the subject.
In addition, PSSs can locate multiple targets with a single measurement if the targets emit in similar frequency bands.
Unfortunately, the RRM algorithms typically do not take this into account.
The PSSRM algorithms presented in this section do not account for different frequency bands observed by the receiver and are always optimized by GAs, likely negatively affecting their runtime.
It is important to note that, unlike our newly proposed algorithm, none of the other algorithms optimize the observed frequency bands to improve resource management.



\section{ResourceTune Algorithm}
\label{sec:methodology}

The core idea of the ResourceTune algorithm is that the receivers can realize multiple tracks and surveys during a single time step if these tasks have similar frequency bands.
This reduces resource consumption and increases efficiency.
Algorithm~\ref{alg:resource_tune} outlines the main steps of the algorithm, and each of the underlying concepts is explained in detail in its subsection.
Each subsection also includes an analysis of the computational complexity of the corresponding step.

\begin{algorithm}
    \caption{ResourceTune}
    \label{alg:resource_tune}
    \KwInput{
        tracks $\tracks$, surveys $\surveys$, shapes $\shapes$, time steps $\timesteps$
    }
    \DontPrintSemicolon

    Task preprocessing (Section~\ref{subsec:preprocessing})\;

    \While{tasks do not change}
    {
        Compute insertion rates of configurations (Section~\ref{subsec:lp})\;

        Construct a tuning plan and send it to the PSS (Section~\ref{subsec:tuning_plan})\;

        Update of historical and current observation rates of tracks and sub-surveys (Section~\ref{subsec:history})\;
    }
\end{algorithm}

\subsection{Task Preprocessing}
\label{subsec:preprocessing}

Task preprocessing is performed before the main loop of the algorithm begins.
It consists of two steps, which are described in the following subsections.

\subsubsection{Survey splitting}

The first step of preprocessing is to split each survey $\survey_\surveyIdx$ into multiple sub-surveys.
This is done because the frequency band of the survey is usually too long for a receiver configuration to observe.
\emph{Sub-survey} $\subsurvey_{\surveyIdx, \subsurveyIdx}$, created by splitting survey $\survey_\surveyIdx$, has a frequency band represented by single-interval $\freq_{\subsurvey_{\surveyIdx, \subsurveyIdx}}$ and goal observation rate $\goalObservation_{\subsurvey_{\surveyIdx, \subsurveyIdx}} \in [0, 1]$, which indicates how often the sub-survey should be realized.
Sub-survey $\subsurvey_{\surveyIdx, \subsurveyIdx}$ is \emph{observed} at a given time if at least one receiver on any sensor node completely observes $\freq_{\subsurvey_{\surveyIdx, \subsurveyIdx}}$ at that time.
Similar to the other tasks, sub-surveys can be observed only \emph{once per time step}.

The process of splitting surveys is described in Algorithm~\ref{alg:sub_survey_construction}.
Frequency band $\freq_{\survey_\surveyIdx}$ of each survey $\survey_\surveyIdx$ is divided from left to right into frequency bands of size $\splitSize$, which is a parameter of the ResourceTune algorithm called \emph{split size}.
Each of these bands corresponds to $\freq_{\subsurvey_{\surveyIdx, \subsurveyIdx}}$ of a newly constructed sub-survey $\subsurvey_{\surveyIdx, \subsurveyIdx}$, whose $\goalObservation_{\subsurvey_{\surveyIdx, \subsurveyIdx}}$ is set to $\goalObservation_{\survey_\surveyIdx}$.
The set of sub-surveys constructed from $\survey_\surveyIdx$ is denoted by $\subsurveys_{\surveyIdx}$.
Similarly, $\subsurveys$ is the set of all sub-surveys.

\begin{algorithm}
    \caption{Survey splitting}
    \label{alg:sub_survey_construction}
    \DontPrintSemicolon
    \KwInput{
        surveys $\surveys$
    }
    \KwOutput{
        sub-surveys $\subsurveys$
    }
    \BlankLine

    \ForAll{$\survey_\surveyIdx \in \surveys$}{
    
        $\subsurveys_{\surveyIdx} \leftarrow \{\}$\;

        \ForAll{$\subsurveyIdx \in \left\{1, \dots, \left\lceil \frac{|\freq_{\survey_\surveyIdx}|}{\splitSize} \right\rceil\right\}$}{

        $\subsurvey_{\surveyIdx, \subsurveyIdx} \leftarrow$ Create new sub-survey such that $\freq_{\subsurvey_{\surveyIdx, \subsurveyIdx}} \leftarrow [\freqLow_{\survey_\surveyIdx} + (\subsurveyIdx-1) \splitSize, \min\{\freqLow_{\survey_\surveyIdx} + \subsurveyIdx \splitSize, \freqUp_{\survey_\surveyIdx}\}]$ and $\goalObservation_{\subsurvey_{\surveyIdx, \subsurveyIdx}} \leftarrow \goalObservation_{\survey_\surveyIdx}$\;

            Insert $\subsurvey_{\surveyIdx, \subsurveyIdx}$ to $\subsurveys_\surveyIdx$\;
        }
    }
\end{algorithm}

Note that by observing all of the sub-surveys from $\subsurveys_\surveyIdx$ as often as stated by their goal observation rates, each frequency of the original survey request $\frequency \in \freq_{\survey_\surveyIdx}$ is also observed as much as demanded.
Therefore, the rest of ResourceTune can only consider the sub-surveys instead of considering each frequency from each survey separately.
Additionally, the sub-surveys are almost identical to the tracks, allowing us to occasionally observe both simultaneously.
The only difference is that the tracks must be observed by at least one receiver on each sensor node, while the sub-surveys only need one receiver on any sensor node.

\subsubsection{Configuration Construction Using Left-right Heuristic}

The second step of preprocessing is to construct configurations $\configurations$.
\emph{Configuration} $\configuration \in \configurations$ is a multiple-interval from $\intervals[\shapes]$.
Configurations can be inserted into tuning plan $\planIdx$ at time step $\timestepIdx$.
This process depends on configuration's weight $\weight_{\configuration} \in \{1, 4\}$. 
If $\weight_{\configuration}$ equals $1$, $\configuration$ is assigned to one receiver $\receiverIdx$ on some sensor node $\nodeIdx$ such that $\plan_{\nodeIdx, \receiverIdx}^{\planIdx, \timestepIdx} \leftarrow \configuration$.
Otherwise, if $\weight_{\configuration}$ equals $4$, $\configuration$ is assigned to one receiver on each sensor node $\receiverIdx, \receiverIdx', \receiverIdx'', \receiverIdx''' \in \receivers$ such that $\plan_{1, \receiverIdx}^{\planIdx, \timestepIdx}, \plan_{2, \receiverIdx'}^{\planIdx, \timestepIdx}, \plan_{3, \receiverIdx''}^{\planIdx, \timestepIdx}, \plan_{4, \receiverIdx'''}^{\planIdx, \timestepIdx} \leftarrow \configuration$. 
In either case, no configuration must be inserted into the concerned receivers before.
Note that weight $\weight_\configuration$ corresponds to the number of receivers that configuration $\configuration$ occupies when inserted into a tuning plan.
See \figurename~\ref{fig:problem_statement} for examples of these insertions.
We say that the configuration observes a track or sub-survey when its insertion into tuning plan makes the object observed.
Note that configurations with weight set to $4$ can realize both tracks and sub-surveys, while those with weight equal to $1$ can realize only sub-surveys.  

The configurations are constructed by the \emph{left-right heuristic}, which allows us to select promising configurations from a vast set of possibilities.
We will collectively refer to each emitter of each track and each sub-survey as the \emph{parent} of a configuration.
The heuristic produces configurations such that each parent is observed by at least one of them, and each configuration realizes as many tasks as possible.
This is described by Algorithm~\ref{alg:block_generation}.

\begin{algorithm}
    \caption{Configuration construction using left-right heuristic}
    \label{alg:block_generation}
    \DontPrintSemicolon
    \KwInput{
        tracks $\tracks$, surveys $\surveys$, sub-surveys $\subsurveys$, shapes $\shapes$
    }
    \KwOutput{
        configurations $\configurations$
    }
    \BlankLine

    $\configurations', \configurations'' \leftarrow \{\}$\;

    \ForAll{$\task \in \left( \bigcup_{\trackIdx = 1}^{|\tracks|} \emitters_\trackIdx \right) \cup \left( \bigcup_{\surveyIdx = 1}^{|\surveys|} \subsurveys_\surveyIdx \right)$}{
    
        $\shape \leftarrow \arg \max_{\shape' \in \shapes} \sum_{y = 1}^{\left\lceil |\shape'| / 2 \right\rceil} \shape'_{2y-1}$ s.t.: $\forall y \in \{1, \dots, \left\lceil |\shape'| / 2 \right\rceil\}: |\task| \leq \shape'_{2y-1} \leq \maxBandwidth_{\task} $\;

        \ForAll{$y \in \{1, \dots, \left\lceil |\shape| / 2 \right\rceil\}$}{

            $\configuration^l \leftarrow$ Left-most configuration such that $\intervalShape(\configuration^l) = \shape$, $\task$ is observed by its $(2y-1)$-th single-interval and $\weight_{\configuration^{l}} \leftarrow 4$\;

            $\configuration^r \leftarrow$ Right-most configuration such that $\intervalShape(\configuration^r) = \shape$, $\task$ is observed by its $(2y-1)$-th single-interval and $\weight_{\configuration^{r}} \leftarrow 4$\;

            Insert $\configuration^l, \configuration^r$ to $\configurations'$\;
        }
    }

    \ForAll{$\configuration \in \configurations'$}{

        $\configuration' \leftarrow $ Make copy of $\configuration$\;

        $\weight_{\configuration'} \leftarrow 1$
        
        Insert $\configuration'$ to $\configurations''$\;            
    }

    $\configurations \leftarrow \configurations' \cup \configurations''$
    
\end{algorithm}

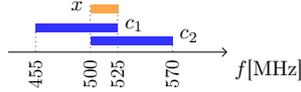
\begin{figure}
    \centering
    \scalebox{0.72}{
    \begin{tikzpicture}[scale=1.0]
        \draw[-{Straight Barb}] (0.5,-0.55)--(4.5,-0.55) node[anchor=north west]{$f$[\si{MHz}]};

        \draw[dotted] (1.0, -0.1)--(1.0, -0.55);
        \node[anchor=east, rotate=90] at (1.0, -0.55) {\small $455$};
        \draw[dotted] (2.0, 0.3)--(2.0, -0.55);
        \node[anchor=east, rotate=90] at (2.0, -0.55) {\small $500$};
        \draw[dotted] (2.5, 0.3)--(2.5, -0.55);
        \node[anchor=east, rotate=90] at (2.5, -0.55) {\small $525$};
        \draw[dotted] (3.5, -0.35)--(3.5, -0.55);
        \node[anchor=east, rotate=90] at (3.5, -0.55) {\small $570$};
    
        \node[anchor=east] at (2, 0.3) {$\task$};
        \node[rectangle, fill=orange!70, minimum height=1.5mm, minimum width=5mm, inner sep=0mm, anchor=west] at (2, 0.25) () {};
        
        \node[rectangle, fill=blue!80, minimum height=1.5mm, minimum width=15mm, inner sep=0mm, anchor=west] at (1.0,-0.1) () {};
        \node[anchor=west] at (2.5,-0.05) {$\configuration_1$};
        \node[rectangle, fill=blue!80, minimum height=1.5mm, minimum width=15mm, inner sep=0mm, anchor=west] at (2.0,-0.35) () {};
        \node[anchor=west] at (3.5,-0.25) {$\configuration_2$};
    \end{tikzpicture}
    }
    \caption{
        Configurations (blue rectangles) that are generated for parent $\task$ (orange rectangle), $\freq_\task = [500, 525]$, and selected shape $\shape = (70)$.
    }
    \label{fig:blocks_1a}
\end{figure}

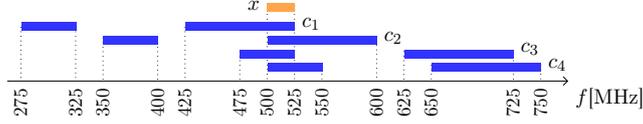
\begin{figure}
    \centering
    \scalebox{0.72}{
    \begin{tikzpicture}[scale=1.0]
        \draw[-{Straight Barb}] (-2.75,-1.1)--(7.5,-1.1) node[anchor=north west]{$f$[\si{MHz}]};

        \draw[dotted] (-2.5, -0.1)--(-2.5, -1.1);
        \node[anchor=east, rotate=90] at (-2.5, -1.1) {\small $275$};
        \draw[dotted] (-1.5, -0.1)--(-1.5, -1.1);
        \node[anchor=east, rotate=90] at (-1.5, -1.1) {\small $325$};
        \draw[dotted] (-1.0, -0.3)--(-1.0, -1.1);
        \node[anchor=east, rotate=90] at (-1.0, -1.1) {\small $350$};
        \draw[dotted] (0.0, -0.3)--(0.0, -1.1);
        \node[anchor=east, rotate=90] at (0.0, -1.1) {\small $400$};
        \draw[dotted] (0.5, -0.1)--(0.5, -1.1);
        \node[anchor=east, rotate=90] at (0.5, -1.1) {\small $425$};
        \draw[dotted] (1.5, -0.55)--(1.5, -1.1);
        \node[anchor=east, rotate=90] at (1.5, -1.1) {\small $475$};
        \draw[dotted] (2.0, 0.3)--(2.0, -1.1);
        \node[anchor=east, rotate=90] at (2.0, -1.1) {\small $500$};
        \draw[dotted] (2.5, 0.3)--(2.5, -1.1);
        \node[anchor=east, rotate=90] at (2.5, -1.1) {\small $525$};
        \draw[dotted] (3, -0.8)--(3, -1.1);
        \node[anchor=east, rotate=90] at (3, -1.1) {\small $550$};
        \draw[dotted] (4.0, -0.35)--(4.0, -1.1);
        \node[anchor=east, rotate=90] at (4.0, -1.1) {\small $600$};
        \draw[dotted] (4.5, -0.55)--(4.5, -1.1);
        \node[anchor=east, rotate=90] at (4.5, -1.1) {\small $625$};
        \draw[dotted] (5.0, -0.8)--(5.0, -1.1);
        \node[anchor=east, rotate=90] at (5.0, -1.1) {\small $650$};
        \draw[dotted] (6.5, -0.55)--(6.5, -1.1);
        \node[anchor=east, rotate=90] at (6.5, -1.1) {\small $725$};
        \draw[dotted] (7.0, -0.8)--(7.0, -1.1);
        \node[anchor=east, rotate=90] at (7.0, -1.1) {\small $750$};
        
        \node[rectangle, fill=orange!70, minimum height=1.5mm, minimum width=5mm, inner sep=0mm, anchor=west] at (2, 0.25) () {};
        \node[anchor=east] at (2, 0.3) {$\task$};
        
        \node[rectangle, fill=blue!80, minimum height=1.5mm, minimum width=20mm, inner sep=0mm, anchor=west] at (0.5,-0.1) () {};
        \node[rectangle, fill=blue!80, minimum height=1.5mm, minimum width=10mm, inner sep=0mm, anchor=west] at (-2.5,-0.1) () {};
        \node[anchor=west] at (2.5,-0.05) {$\configuration_1$};

        \node[rectangle, fill=blue!80, minimum height=1.5mm, minimum width=20mm, inner sep=0mm, anchor=west] at (2.0,-0.35) () {};
        \node[rectangle, fill=blue!80, minimum height=1.5mm, minimum width=10mm, inner sep=0mm, anchor=west] at (-1.0,-0.35) () {};
        \node[anchor=west] at (4,-0.3) {$\configuration_2$};
        
        \node[rectangle, fill=blue!80, minimum height=1.5mm, minimum width=10mm, inner sep=0mm, anchor=west] at (1.5,-0.6) () {};
        \node[rectangle, fill=blue!80, minimum height=1.5mm, minimum width=20mm, inner sep=0mm, anchor=west] at (4.5,-0.6) () {};
        \node[anchor=west] at (6.5,-0.5) {$\configuration_3$};
        
        \node[rectangle, fill=blue!80, minimum height=1.5mm, minimum width=10mm, inner sep=0mm, anchor=west] at (2.0,-0.85) () {};
        \node[rectangle, fill=blue!80, minimum height=1.5mm, minimum width=20mm, inner sep=0mm, anchor=west] at (5.0,-0.85) () {};
        \node[anchor=west] at (7.0,-0.8) {$\configuration_4$};
    \end{tikzpicture}
    }
    \caption{
        Configurations (blue rectangles) that are generated for parent $\task$ (orange rectangle), $\freq_\task = [500, 525]$, and selected shape $\shape = (50, 100, 100)$.
    }
    \label{fig:blocks_1b}
\end{figure}

For each parent $\task$ the heuristic selects an appropriate shape $\shape \in \shapes$ such that each of its single-intervals is wide enough to realize $\task$ and the total width of observed frequencies is maximized.
If $\task$ is an emitter, all single-intervals of $\shape$ must also have a width less than or equal to $\maxBandwidth_{\task}$.
Then, for each single-interval of $\shape$, the heuristic generates two configurations such that both realize $\task$, one positioned as far left as possible $\configuration^l$ and the other as far right as possible $\configuration^r$.
\figurename~\ref{fig:blocks_1a} shows two configurations constructed by the left-right heuristic for parent $\task$ when the selected shape is $(70)$.
The same is shown for shape $(50, 100, 100)$ in \figurename~\ref{fig:blocks_1b}.
The number of configurations constructed for $\task$ is two times the number of single-intervals in $\shape$.

Each configuration is then duplicated, setting the weight of the original to $4$ and the weight of the duplicate to $1$.
This duplication is intended primarily to conserve resources. 
In some cases, it may be optimal to use this configuration to observe only sub-surveys, so it is unnecessary to occupy a receiver on each sensor node.

\subsubsection{Computational Complexity}

The worst-case time complexity of the preprocessing step, denoted $\complexity_{1}(\tracks, \surveys, \splitSize)$, corresponds to the number of constructed configurations.
This number is linearly dependent on the number of parents $\parentCount(\tracks, \surveys, \splitSize)$, which is equal to the total number of emitters plus the number of sub-surveys:
\begin{equation}
    \parentCount(\tracks, \surveys, \splitSize) = \sum_{\trackIdx = 1}^{|\tracks|} |\emitters_\trackIdx| + \sum_{\survey_\surveyIdx \in \surveys} \left\lceil \frac{|\freq_{\survey_\surveyIdx}|}{\splitSize} \right\rceil.
\end{equation}
Consequently,
\begin{equation}
    \complexity_{1}(\tracks, \surveys, \splitSize) \in \mathcal{O}\left(\parentCount(\tracks, \surveys, \splitSize)\right).
\end{equation}

\subsection{Computing Goal Insertion Rate of Configurations}
\label{subsec:lp}

During $\planIdx$-th iteration of the algorithm, each configuration $\configuration$ has \emph{goal insertion rate} $\goalInsertion_\configuration^\planIdx \in \mathbb{R}_{\geq 0}$ that specifies how many times it should be inserted into a tuning plan.
First, the set of all configurations $\configurations$ is filtered, creating $\configurations'$ so that each remaining configuration observes a unique set of tracks and sub-surveys.

The algorithm maintains information about how often each track and sub-survey $\task \in (\tracks \cup \subsurveys)$ should be observed in the tuning plan constructed during the algorithm's $\planIdx$-th iteration.
This information is called the \emph{current observation rate} $\currentObservation_{\task}^\planIdx$. 
During the first iteration, $\currentObservation_{\task}^1$ is initialized to goal measurement rate $\goalObservation_\task$. 
Section~\ref{subsec:history} describes in detail how the value of $\currentObservation_{\task}^\planIdx$ is computed in later iterations of the algorithm.

Now that we have all the necessary information, the algorithm determines the optimal goal insertion rates of the configurations from $\configurations'$ by solving an LP described by Equations~\eqref{eq:lp_objective}--\eqref{eq:lp_reals}.
The objective is to minimize the use of receivers \eqref{eq:lp_objective} while setting the goal insertion rates of the configurations so that each track \eqref{eq:lp_tracks} and sub-survey \eqref{eq:lp_sub_surveys} is observed by these configurations as often as demanded by their current observation rate.
\begin{alignat}{1}
    \min_{
        \goalInsertion^\planIdx
    } & \sum_{\configuration \in \configurations'}  \weight_\configuration \, \goalInsertion_\configuration^\planIdx \label{eq:lp_objective} 
\end{alignat}
subject to:
{\allowdisplaybreaks
\begin{alignat}{3}
    \sum_{\configuration \in \configurations': \, \configuration \text{ observes } \track_\trackIdx} \goalInsertion_\configuration^\planIdx &\geq \currentObservation_{\track_\trackIdx}^\planIdx && \qquad \forall \track_\trackIdx \in \tracks \label{eq:lp_tracks}
    \\
    \sum_{\configuration \in \configurations': \, \configuration \text{ observes } \subsurvey_{\surveyIdx, \subsurveyIdx}} \goalInsertion_\configuration^\planIdx &\geq \currentObservation_{\subsurvey_{\surveyIdx, \subsurveyIdx}}^\planIdx && \qquad \forall \subsurvey_{\surveyIdx, \subsurveyIdx} \in \subsurveys_\surveyIdx \; \forall \survey_\surveyIdx \in \surveys \label{eq:lp_sub_surveys}
    \\
    \goalInsertion_\configuration^\planIdx &\in \mathbb{R}_{\geq 0} && \qquad \forall \configuration \in \configurations'. \label{eq:lp_reals}
\end{alignat}
}The goal insertion rates of the configurations that are not in $\configurations'$, i.e., those that were filtered out in the first step, are set to 0.
Consequently, inserting each configuration $\configuration \in \configurations$ into tuning plan $\plan^\planIdx$ at its goal insertion rate $\goalInsertion_{\configuration}^{\planIdx}$ ensures that all tasks are fully observed.
For this reason, we minimize the use of receivers to maximize the likelihood that all of the configurations will be inserted into the tuning plan as often as needed.

\subsubsection{Computational Complexity}

The asymptotic time complexity of this step is dominated by solving the LP.
\citet{vaidya_lp} proved that the time complexity of solving an LP is bounded~by:
\begin{equation}
    \mathcal{O}((\variableCount + \constrainCount)^{1.5} \constrainCount \precision),
\end{equation}
where $\variableCount$ is the number of variables, $\constrainCount$ is the number of constraints, and $\precision$ is the bit precision.
In our case, $\variableCount$ linearly depends on $\parentCount(\tracks, \surveys, \splitSize)$ and $\constrainCount$ corresponds to $|\tracks| + \sum_{\survey_\surveyIdx \in \surveys} \left\lceil \frac{|f_{\survey_\surveyIdx}|}{\splitSize} \right\rceil$.
Therefore, the time complexity of this step $\complexity_2(\tracks, \surveys, \splitSize)$ can be written as:
\begin{equation}
    \label{eq:t3_large}
    \begin{split}
        \complexity_2(\tracks, \surveys, \splitSize) \in \mathcal{O}\left(\left(|\tracks| + \sum_{\survey_\surveyIdx \in \surveys} \left\lceil \frac{|f_{\survey_\surveyIdx}|}{\splitSize} \right\rceil + \parentCount(\tracks, \surveys, \splitSize)\right)^{1.5} \cdot \parentCount(\tracks, \surveys, \splitSize)\right).
    \end{split}
\end{equation}
Since we know that each track request has at least one emitter,
\begin{equation}
    |\tracks| \leq \sum_{\trackIdx = 1}^{|\tracks|} |\emitters_\trackIdx|,
\end{equation}
we can substitute $|\tracks|$ with $\sum_{\trackIdx = 1}^{|\tracks|} |\emitters_\trackIdx|$. 
Then \eqref{eq:t3_large} can be simplified:
\begin{equation}
    \complexity_2(\tracks, \surveys, \splitSize) \in \mathcal{O}(\parentCount^{2.5}(\tracks, \surveys, \splitSize)).
\end{equation}

\subsection{Construction of Tuning Plan}
\label{subsec:tuning_plan}

ResourceTune attempts to construct a tuning plan that respects the goal insertion rates of all configurations.
However, due to the limited number of receivers and potentially large number of tasks, this may be impossible.
The entire construction process of plan $\plan^\planIdx$ is described by Algorithm~\ref{alg:tuning_plan_construction}.
First, empty tuning plan $\plan^\planIdx$ with $|\timesteps|$ time steps and lexicographic queue $\queue$ containing configurations with goal insertion rates greater than zero are initialized.
The queue maintains the configurations in increasing lexicographic order according to triplet $(\realizedObservation_{\configuration}^\planIdx, -\weight_\configuration, -\goalInsertion_\configuration^\planIdx)$, where $\realizedObservation_{\configuration}^\planIdx$ denotes the proportion of time steps in which configuration $\configuration$ is inserted into tuning plan $\plan^\planIdx$.
This means that the queue sorts the configurations from the least to the most inserted. 
Then, configurations with $\weight_\configuration$ equal to $4$ are preferred because they are more difficult to insert into the tuning plan. 
Finally, configurations with a higher insertion rate come before those with a lower insertion rate.
Until the queue is empty, the algorithm removes the first configuration $\configuration$ from the queue and attempts to insert it into $\plan^\planIdx$ twice, as shown at Lines~\ref{alg:tuning_plan_construction:insertion_start}--\ref{alg:tuning_plan_construction:insertion_end}. 
The first attempt, described at Lines~\ref{alg:tuning_plan_construction:insertion_start}--\ref{alg:tuning_plan_construction:p1_end}, first generates the set of all positions $\positions_1$ where $\configuration$ can be inserted into $\plan^\planIdx$ and each position $\position \in \positions_1$ satisfies the following conditions w.r.t. $\configuration:$ no overlap and no fragmentation.

The \emph{no overlap} condition means that it is not possible to insert $\configuration$ into $\plan^\planIdx$ at time step $\timestepIdx$ if a configuration already inserted there observes at least one track or sub-survey that is also observed by $\configuration$.
This is done to prevent the unnecessary repetition of configurations, since each track or sub-survey can only be observed once per time step.
The \emph{no fragmentation} condition helps to create a compact tuning plan by filtering out positions where inserting $\configuration$ would decrease the \emph{plan cohesion}, which is defined as the number of configurations with weight equal to $4$ that can be inserted into $\plan^\planIdx$.
As a consequence of this condition, the configurations with weight equal to $1$ tend to cluster at the same time steps, resulting in a plan with more positions for future insertion of configurations with weight set to $4$. 

If $\positions_1$ is not empty, one of the positions $\position \in \positions_1$ is selected, and $\configuration$ is inserted into $\plan^\planIdx$ at this position, as described in Algorithm~\ref{alg:insert_block}.
Then if goal insertion rate $\goalInsertion_\configuration^\planIdx$ is still larger than $\realizedObservation_{\configuration}^\planIdx$, the configuration is returned to the queue.
If the first insertion attempt fails because there are no feasible positions, the second insertion attempt, which is less constrained than the first, is made as described in Algorithm~\ref{alg:tuning_plan_construction} at Lines~\ref{alg:tuning_plan_construction:p2_start}--\ref{alg:tuning_plan_construction:insertion_end}.
If even the second attempt fails, the configuration is not returned to the queue.
These configuration insertions are repeated until the queue is empty.

\begin{algorithm}
    \caption{Construction of tuning plan $\plan^\planIdx$}
    \label{alg:tuning_plan_construction}
    \DontPrintSemicolon
    \SetKwFunction{InsertConfiguration}{InsertConfiguration}
    \KwInput{
        configurations $\configurations$
    }
    \KwOutput{tuning plan $\plan^\planIdx$}
    \BlankLine
    Initialize empty tuning plan $\plan^\planIdx$ with $|\timesteps|$ time steps\;
    
    Initialize empty lexicographic queue $\queue$\;
    
    \ForAll{$\configuration \in \configurations$}
    {
        \If{$\goalInsertion_\configuration^\planIdx > 0$}
        {
            Insert configuration $\configuration$ to $\queue$ w.r.t. $(\realizedObservation_{\configuration}^\planIdx, -\weight_\configuration, -\goalInsertion_\configuration^\planIdx)$\;
        }
    }
    \While{$\queue$ is not empty}
    {
        $\configuration \xleftarrow{}$ Pop configuration from $\queue$\; \label{alg:tuning_plan_construction:pop}

        $\positions_1 \leftarrow \{\position \subset \plan^\planIdx \;|\; \position \text{ is for } \configuration: \text{no overlap; } \text{no fragmentation}\}$\; \label{alg:tuning_plan_construction:insertion_start}

        \If{$\positions_1$ is not empty}
        {
            $\plan^\planIdx, \queue \leftarrow$ \InsertConfiguration{$\plan^\planIdx$, $\queue$, $\configuration$, $\positions_1$}\;
            
            \textbf{continue}\; \label{alg:tuning_plan_construction:p1_end}
        }
        
        $\positions_2 \leftarrow \{\position \subset \plan^\planIdx \;|\; \position \text{ is for } \configuration: \text{no overlap}\}$\; \label{alg:tuning_plan_construction:p2_start}

        \If{$\positions_2$ is not empty}
        {
            $\plan^\planIdx, \queue \leftarrow$ \InsertConfiguration{$\plan^\planIdx$, $\queue$, $\configuration$, $\positions_2$}\;
            
            \textbf{continue}\; \label{alg:tuning_plan_construction:insertion_end}
        }
    }
\end{algorithm}

\begin{algorithm}
    \caption{InsertConfiguration}
    \label{alg:insert_block}
    \DontPrintSemicolon
    \KwInput{
        tuning plan $\plan^\planIdx$, 
        priority queue $\queue$,
        configuration $\configuration$,
        set of positions $\positions$
    }
    \KwOutput{
        tuning plan $\plan^\planIdx$, 
        priority queue $\queue$
    }
    \BlankLine

    $\position \leftarrow$ Select position from $\positions$\;

    Insert configuration $\configuration$ to $\plan^\planIdx$ at position $\position$\;

    \If{$\goalInsertion_\configuration^\planIdx > \realizedObservation_{\configuration}^\planIdx$}
    {
        Insert configuration $\configuration$ to $\queue$ w.r.t. $(\realizedObservation_{\configuration}^\planIdx, -\weight_\configuration, -\goalInsertion_\configuration^\planIdx)$\;
    }
\end{algorithm}

\subsubsection{Computational Complexity}

The time complexity of constructing the queue depends on the number of configurations and is therefore equal to $\mathcal{O}(\parentCount(\tracks, \surveys, \splitSize))$.
The number of iterations during plan construction does not depend on the number of tasks, but on the number of time steps $|\timesteps|$ and the number of receivers, which is constant.
The time complexity of each iteration is the sum of two factors.
The first factor is the complexity of generating position sets $\positions_1$ and $\positions_2$, which depends on the number of time steps $|\timesteps|$ and the number of receivers.
The second factor is the complexity of priority queue management, which is $\mathcal{O}\left(\log\left(\parentCount(\tracks, \surveys, \splitSize)\right)\right)$.
Therefore, the asymptotic time complexity of one iteration is
\begin{equation}
    \mathcal{O}\left(\parentCount(\tracks, \surveys, \splitSize) + |\timesteps| \left(|\timesteps| + \log\parentCount(\tracks, \surveys, \splitSize)\right)\right). 
\end{equation}
For PSS, we can assume that $|\timesteps|$ is a small constant, and consequently,
\begin{equation}
    \complexity_3(\tracks, \surveys, \splitSize) \in \mathcal{O}\left(\parentCount(\tracks, \surveys, \splitSize)\right).
\end{equation}

\subsection{Update of Historical Observation Rate}
\label{subsec:history}

During the runtime of the algorithm, tracks or sub-surveys may not be observed enough. 
This is likely to happen when there are too many tasks that have high goal observation rates.
On the other hand, they may sometimes be observed more often than necessary.
This can happen when a configuration observes two tracks with different goal observation rates. 
As a result, the track with the lower observation rate is observed more often than necessary.
Therefore, including information about previous observations in the current observation rate is useful for balancing these deviations.

\emph{Historical observation rate} $\historicalObservation_\task^\planIdx$ of track or sub-survey $\task \in (\tracks \cup \subsurveys)$ at tuning plan $\planIdx$ is defined:
\begin{equation}
    \begin{split}
    \historicalObservation_\task^\planIdx = \sum_{\planIdx' = 1}^{\planIdx} \discount^{\planIdx-\planIdx'} \realizedObservation_{x}^{\planIdx'} = \discount \historicalObservation_\task^{\planIdx - 1} + \realizedObservation_{x}^\planIdx,    
    \end{split}
\end{equation}
where $\discount \in (0, 1)$.
The $\discount$ is a parameter of the algorithm called the \emph{discount factor}.
It influences how much the previous tuning plans influence the current observation rate. 
The higher the $\discount$, the more influence the previous plans have.

The algorithm's goal for the $(\planIdx+1)$-th iteration is to realize $\task$ so many times that $\historicalObservation_\task^{\planIdx + 1}$ divided by the sum of a geometric series with the common ratio $\discount$ is greater than or equal to~$\goalObservation_\task$:
\begin{equation}
    \goalObservation_\task 
    \leq \frac{\gamma^{\planIdx} \realizedObservation_{\task}^1 + \dots + \discount \realizedObservation_{\task}^\planIdx + \realizedObservation_{\task}^{\planIdx + 1}}{\discount^\planIdx + \dots + \discount + 1}
    = \frac{\gamma^{\planIdx} \realizedObservation_{\task}^1 + \dots + \discount \realizedObservation_{\task}^\planIdx + \realizedObservation_{\task}^{\planIdx + 1}}{\frac{1-\discount^{\planIdx+1}}{1-\discount}}.
\end{equation}
For this to be true, we should set $\currentObservation_\task^{\planIdx + 1}$ equal to $\realizedObservation_{\task}^{\planIdx + 1}$. 
Now, we can simplify and rearrange the inequality to solve for $\currentObservation_\task^{\planIdx + 1}$:
\begin{alignat}{1}
    \goalObservation_\task &\leq \frac{\discount \historicalObservation_\task^\planIdx + \currentObservation_\task^{\planIdx + 1}}{\frac{1-\discount^{\planIdx+1}}{1-\discount}}
    \\
    \currentObservation_\task^{\planIdx + 1} &\geq \frac{1-\discount^{\planIdx+1}}{1-\discount} \goalObservation_\task - \gamma \historicalObservation_\task^\planIdx.
\end{alignat}
Therefore, setting $\currentObservation_\task^{\planIdx + 1}$ equal to $\frac{1-\discount^{\planIdx+1}}{1-\discount} \goalObservation_\task - \gamma \historicalObservation_\task^\planIdx$ ensures that $\task$ is observed with the goal observation rate $\goalObservation_\task$ throughout its existence.

\subsubsection{Computational Complexity}

The above computation must be performed for each track request and sub-survey, so the asymptotic time complexity of this step is
\begin{equation}
    \complexity_4(\tracks, \surveys, \splitSize) \in \mathcal{O}(\parentCount(\tracks, \surveys, \splitSize)).
\end{equation}

\subsection{Overall Computational Complexity}
\label{subsec:complexity}

We have provided the worst-case time complexity for each step of ResourceTune.
The complexities $\complexity_2$, $\complexity_3$, and $\complexity_4$ can be aggregated, creating the overall asymptotic complexity of one tuning plan construction
\begin{equation}
    \complexity(\tracks, \surveys, \splitSize) \in \mathcal{O}(\parentCount^{2.5}(\tracks, \surveys, \splitSize)),
\end{equation}
which corresponds to the time complexity of solving the LP.
As can be seen, the proposed algorithm has a low computational time complexity that is polynomial.
Therefore, assuming that $\splitSize$ is selected reasonably, we are confident that the proposed algorithm works in real time.


\section{Experimental Evaluation}
\label{sec:experiments}

This section compares the ResourceTune algorithm with other methods.
First, we describe our experimental setup and how our problem instances were generated.
Then, we focus on tuning the parameters of ResourceTune, primarily the split size $\splitSize$.
We also demonstrate the advantages of using a left-right heuristic for configuration construction.
Finally, we compare ResourceTune with two algorithms.
The first algorithm, called greedy, constructs tuning plans using a greedy approach. It is inspired by a typical radar resource management algorithm called the time balancing algorithm \citep{butler_time_balancing}.
The second algorithm, denoted by GA, is a genetic algorithm used to construct tuning plans, as described by \citet{suja_scheduling}. 
It is the state of the art in PSSRM.
Further description of both algorithms can be found in \ref{app:algorithms}.

\subsection{Problem Instances and Experimental Setup}

In the experiments, we consider a PSS whose tuning plan duration is $2$ seconds, each plan consists of $|\timesteps| = 10$ time steps, and receiver shapes are $\shapes = \{(y) \,|\, y \in \mathbb{N}: 10 \leq y \leq 100\} \cup \{(100, 100, 100)\}$.
This means that each evaluated method had two seconds to construct a tuning plan.
The experimental evaluation was carried out using randomly generated problem instances.

In each instance, there were $50$ tracks, each with a random number of emitters determined by uniform sampling between $1$ and $3$.
The frequency band $\freq_{\emitter_{\trackIdx, \emitterIdx}}$ of each emitter $\emitter_{\trackIdx, \emitterIdx}$ was sampled uniformly from single-interval $[10\,000, 16\,000]$ such that $|\freq_{\emitter_{\trackIdx, \emitterIdx}}|$ was between $1$ and $50$.
In addition, each of the emitters had its maximum bandwidth $\maxBandwidth_{\emitter_{\trackIdx, \emitterIdx}}$ set to $100$ with a probability of $75\%$.
Otherwise, $\maxBandwidth_{\emitter_{\trackIdx, \emitterIdx}}$ was randomly uniformly sampled between $|\freq_{\emitter_{\trackIdx, \emitterIdx}}|$ and $100$.
Each instance also included $10$ surveys whose frequency bands $\freq_{\survey_\surveyIdx}$ were determined by randomly dividing single-interval $[10\,000, 16\,000]$.

We must describe how the goal observation rates of the tracks and surveys were determined.
To do this, let us consider a simple PSSRM algorithm that observes tracks and surveys separately.
Furthermore, we assume that it only observes one track per configuration insertion into the tuning plan.
This allows us to compute the expected proportion of system resources needed to observe all tracks, denoted by $\utilizationTrack$:
\begin{equation}
    \utilizationTrack = \frac{1}{2} \sum_{\track_\trackIdx \in \tracks} \goalObservation_{\track_\trackIdx},
\end{equation}
which is the sum of the goal observation rates of each track divided by the number of receivers per sensor node, that is, $2$. 
We also assume that the simple algorithm observes the surveys using configurations with shape $(100, 100, 100)$.
Thus, the expected proportion of system resources used to observe all surveys, denoted by $\utilizationSurvey$ is:
\begin{equation}
    \utilizationSurvey = \frac{1}{8} \sum_{\survey_\surveyIdx \in \surveys} \goalObservation_{\survey_\surveyIdx} \left\lceil \frac{|\freq_{\survey_{\surveyIdx}}|}{200} \right\rceil,
\end{equation}
where the result of the ceiling function is the number of receivers needed to observe the survey~$\survey_\surveyIdx$, and the resulting sum is divided by 8 because it is the total number of receivers in the considered PSS.
Consequently, the \emph{expected resource utilization} $\utilization$ is
\begin{equation}
    \label{eq:utilization}
    \utilization = \utilizationTrack + \utilizationSurvey.
\end{equation}

Using this, the goal observation rates of tracks and surveys are determined by first sampling a number from $(0,1]$ for each of these tasks and then normalizing these values so that Equation~\eqref{eq:utilization} and the following two equations hold:
\begin{alignat}{1}
    \utilizationTrack &= \proportion \utilization,
    \\
    \utilizationSurvey &= (1 - \proportion) \utilization,
\end{alignat}
where $\utilization$ and $\proportion$ are scenario parameters.
The parameter $\utilization \in \mathbb{R}_{>0}$ is the expected resource utilization needed to observe all tasks and the parameter $\proportion \in [0, 1]$, called the \emph{track proportion}, determines what proportion of these estimated resources will be allocated to the tracks.
The exact values of these parameters used in the experiments are listed in the relevant subsection.

The algorithms were implemented in Python 3.11, the COIN-OR Branch-and-Cut (CBC) solver \citep{cbc} was used to solve LPs, and the pymoo package \citep{pymoo} was used to implement GA.
All of the experiments were conducted on a computer with an AMD EPYC v4 (3.25 GHz) CPU, 32 GB of RAM, and no dedicated GPU.

\subsection{Parameter Tuning of ResourceTune}

ResourceTune has two parameters whose values must be first determined, discount factor $\discount$ and split size $\splitSize$.
The discount factor only influences how similar historical observation rate~$\historicalObservation_\task^\planIdx$ will be to the average observation rate. 
The closer $\discount$ is to $1.0$, the more similar they are, and ResourceTune's behavior is more exact.
Consequently, $\discount$ should be set to any value close to $1.0$.
In our case, it was set to $0.99999$.

To determine the optimal value of $\splitSize$, we experimented with the following six values: $3$, $4$, $5$, $20$, $50$, $100$.
We generated $50$ problem instances for $\utilization = 2.0$ and each track proportion $\proportion \in \{0.25, 0.50, 0.75\}$.
Then, for each $\splitSize$ value and each problem instance, we let the ResourceTune algorithm construct $100$ consecutive tuning plans, i.e., $|\plans| = 100$.

The results of this experiment are shown in Table~\ref{tab:split_size}.
As the value of $\splitSize$ decreases from $100$ to $5$, the algorithm's runtime increases and the objective value improves.
The observed increase in runtime supports the theoretical conclusions drawn in Section~\ref{subsec:complexity}.
The improvement in objective value shows that smaller sub-surveys are better for maintaining information about surveyed frequencies.
When $\splitSize$ was less than $5$, the objective value began to deteriorate.
This occurred because the LP runtime increased, leaving little time to construct a tuning plan. 
When $\splitSize$ was $3$, the LP runtime was so long that it did not finish in $2$ seconds. 
Based on the results presented, $\splitSize$ was set to $5$ for subsequent experiments.

\begin{table}[]
\centering
\caption{
    Runtimes and objective values of ResourceTune for different values of $\splitSize$ and $\proportion$.
}
\label{tab:split_size}
\begin{tabular}{@{}rrrrrrr@{}}
\toprule
\multicolumn{1}{l}{} & \multicolumn{1}{l}{} & \multicolumn{4}{c}{Runtime [s]} & \multicolumn{1}{l}{} \\ \cmidrule(lr){3-6}
\multicolumn{1}{l}{} & \multicolumn{1}{l}{} & \multicolumn{1}{c}{Linear programming} & \multicolumn{1}{c}{Plan construction} & \multicolumn{2}{c}{Total} & \multicolumn{1}{c}{Objective $\obj$} \\ \cmidrule(l){3-3} \cmidrule(l){4-4} \cmidrule(l){5-6} \cmidrule(l){7-7}
\multicolumn{1}{c}{$\splitSize$} & \multicolumn{1}{c}{$\proportion$} & \multicolumn{1}{c}{Mean $\pm$ SD} & \multicolumn{1}{c}{Mean $\pm$ SD} & \multicolumn{1}{c}{Mean $\pm$ SD} & \multicolumn{1}{c}{Max} & \multicolumn{1}{c}{Mean $\pm$ SD} \\ \midrule
\rowcolor[HTML]{EFEFEF} 
\cellcolor[HTML]{EFEFEF} & 0.25 & \cellcolor[HTML]{EFEFEF}$2.00 \pm 0.00$ & --- & \cellcolor[HTML]{EFEFEF}$2.00 \pm 0.00$ & $2.00$ & $4.65 \pm 0.64$ \\
\rowcolor[HTML]{EFEFEF} 
\cellcolor[HTML]{EFEFEF} & \cellcolor[HTML]{EFEFEF}0.50 & $2.00 \pm 0.00$ & --- & $2.00 \pm 0.00$ & $2.00$ & \cellcolor[HTML]{EFEFEF}$4.43 \pm 0.43$ \\
\rowcolor[HTML]{EFEFEF} 
\multirow{-3}{*}{\cellcolor[HTML]{EFEFEF}3} & \cellcolor[HTML]{EFEFEF}0.75 & $2.00 \pm 0.00$ & --- & $2.00 \pm 0.00$ & $2.00$ & \cellcolor[HTML]{EFEFEF}$4.22 \pm 0.21$ \\
 & 0.25 & $1.56 \pm 0.17$ & $0.22 \pm 0.07$ & $1.78 \pm 0.16$ & $2.00$ & $1.63 \pm 0.48$ \\
 & 0.50 & $1.54 \pm 0.17$ & $0.16 \pm 0.07$ & $1.70 \pm 0.17$ & $2.00$ & $0.90 \pm 0.29$ \\
\multirow{-3}{*}{4} & 0.75 & $1.51 \pm 0.17$ & $0.10 \pm 0.05$ & $1.61 \pm 0.17$ & $2.00$ & $0.32 \pm 0.51$ \\
\rowcolor[HTML]{EFEFEF} 
\cellcolor[HTML]{EFEFEF} & 0.25 & $0.92 \pm 0.11$ & \cellcolor[HTML]{EFEFEF}$0.21 \pm 0.06$ & \cellcolor[HTML]{EFEFEF}$1.13 \pm 0.12$ & $1.61$ & \cellcolor[HTML]{EFEFEF}$\mathbf{1.52 \pm 0.42}$ \\
\rowcolor[HTML]{EFEFEF} 
\cellcolor[HTML]{EFEFEF} & \cellcolor[HTML]{EFEFEF}0.50 & $0.91 \pm 0.10$ & $0.15 \pm 0.06$ & $1.06 \pm 0.12$ & $1.53$ & $\mathbf{0.86 \pm 0.26}$ \\
\rowcolor[HTML]{EFEFEF} 
\multirow{-3}{*}{\cellcolor[HTML]{EFEFEF}5} & \cellcolor[HTML]{EFEFEF}0.75 & $0.89 \pm 0.10$ & $0.09 \pm 0.05$ & $0.98 \pm 0.11$ & $1.40$ & $\mathbf{0.24 \pm 0.15}$ \\
 & 0.25 & $0.12 \pm 0.02$ & $0.15 \pm 0.04$ & $0.27 \pm 0.05$ & $0.52$ & $1.57 \pm 0.44$ \\
 & 0.50 & $0.12 \pm 0.02$ & $0.10 \pm 0.04$ & $0.22 \pm 0.04$ & $0.47$ & $0.89 \pm 0.27$ \\
\multirow{-3}{*}{20} & 0.75 & $0.12 \pm 0.03$ & $0.06 \pm 0.03$ & $0.18 \pm 0.04$ & $0.44$ & $0.25 \pm 0.15$ \\
\rowcolor[HTML]{EFEFEF} 
\cellcolor[HTML]{EFEFEF} & 0.25 & $0.08 \pm 0.01$ & \cellcolor[HTML]{EFEFEF}$0.12 \pm 0.03$ & \cellcolor[HTML]{EFEFEF}$0.20 \pm 0.04$ & $0.45$ & \cellcolor[HTML]{EFEFEF}$1.69 \pm 0.50$ \\
\rowcolor[HTML]{EFEFEF} 
\cellcolor[HTML]{EFEFEF} & \cellcolor[HTML]{EFEFEF}0.50 & $0.08 \pm 0.01$ & $0.08 \pm 0.03$ & $0.16 \pm 0.03$ & $0.35$ & $0.98 \pm 0.31$ \\
\rowcolor[HTML]{EFEFEF} 
\multirow{-3}{*}{\cellcolor[HTML]{EFEFEF}50} & \cellcolor[HTML]{EFEFEF}0.75 & $0.08 \pm 0.02$ & $0.05 \pm 0.02$ & $0.13 \pm 0.03$ & $0.60$ & $0.32 \pm 0.17$ \\
 & 0.25 & $0.06 \pm 0.01$ & $0.10 \pm 0.03$ & $0.16 \pm 0.03$ & $0.46$ & $1.94 \pm 0.55$ \\
 & 0.50 & $0.06 \pm 0.01$ & $0.06 \pm 0.02$ & $0.13 \pm 0.03$ & $0.31$ & $1.21 \pm 0.37$ \\
\multirow{-3}{*}{100} & 0.75 & $0.06 \pm 0.01$ & $0.04 \pm 0.02$ & $0.10 \pm 0.02$ & $0.29$ & $0.46 \pm 0.20$ \\ \bottomrule
\end{tabular}
\end{table}

It is interesting to note the influence of track proportion $\proportion$ on both the runtime and the objective value.
The tuning plan construction runtime is longer when survey tasks have higher observation rates. 
Due to these higher rates, configurations with $\weight_{\configuration}$ equal to $1$ have higher insertion rates while occupying less space in the tuning plan. 
Therefore, more configurations are inserted into a tuning plan than when the track proportion is higher.
Similarly, the objective value is worse when the surveys have higher observation rates.
There are two possible explanations for this phenomenon.
The first explanation is that this difference is caused by an imperfect computation of $\utilizationSurvey$, which underestimates the resources required to observe all surveys.
The second explanation is that it is more difficult to observe multiple sub-surveys with a single configuration than it is to observe multiple tracks.

\subsection{Comparing Configuration Construction Approaches}

To demonstrate the advantages of the left-right heuristic, which is used for configuration construction, we compared its behavior with two alternative approaches.
The first approach is called \emph{centered} and is nearly identical to the left-right heuristic, except the configurations are centered on the parent instead of being shifted left and right.
Consequently, the number of configurations produced using the centered approach is half the number constructed using the left-right heuristic.
The second approach, called \emph{left-center-right}, combines the centered approach and the left-right heuristic.

To compare these three approaches, we generated $50$ instances for each pair of scenario parameters $\utilization \in \{1.0, 2.0, 3.0\}$ and $\proportion \in \{0.25, 0.50, 0.75\}$.
Then, each approach constructed configurations, after which the LP described in Section~\ref{subsec:lp} was constructed and solved.
The average number of configurations, i.e., $|\configurations|$, for the centered, left-right, and left-center-right approaches was $4750.60 \pm 104.94$, $9493.68 \pm 217.04$, and $14242.72 \pm 321.66$, respectively.
Similarly, the average number of unique configurations, i.e., $|\configurations'|$, was $2616.50 \pm 101.95$, $3434.64 \pm 121.92$, and $5168.16 \pm 132.21$, respectively.
These values are consistent with the descriptions of the approaches.

Table~\ref{tab:configurations} reports the number of configurations with an insertion rate greater than 0.0 (in the \#Non-zero column), the LP's runtime, and the LP's objective value for each approach and scenario parameter.
A low number of non-zero configurations is preferred because it simplifies the construction of the tuning plan.
Although the centered approach creates LPs that can be solved quickly, it results in significantly higher objective values and a greater number of non-zero configurations than the other approaches.
By contrast, the left-center-right approach yields the smallest objective values across all scenarios, but its LPs require significantly more time to solve.
The left-right heuristic has the lowest number of non-zero configurations out of all tested approaches across all scenarios.
Furthermore, it can be seen as the ideal compromise between runtime and objective value.
This is because its runtime is often the best, or nearly the best, and the same can be said about its objective values.

\begin{landscape}
\begin{table}[]
\centering
\caption{
    A comparison of three different approaches to generating configurations based on the number of non-zero configurations, runtime and the objective value of the subsequent LP for different values of scenario parameters $\utilization$ and~$\proportion$.
}
\label{tab:configurations}
\begin{tabular}{@{}rrrrrrrrrrr@{}}
\toprule
\multicolumn{1}{c}{} & \multicolumn{1}{c}{} & \multicolumn{3}{c}{Centered} & \multicolumn{3}{c}{Left-right heuristic} & \multicolumn{3}{c}{Left-center-right} \\ \cmidrule(l){3-5} \cmidrule(l){6-8} \cmidrule(l){9-11} 
\multicolumn{1}{c}{$\utilization$} & \multicolumn{1}{c}{$\proportion$} & \multicolumn{1}{c}{\#Non-zero} & \multicolumn{1}{c}{LP time {[}s{]}} & \multicolumn{1}{c}{LP obj.} & \multicolumn{1}{c}{\#Non-zero} & \multicolumn{1}{c}{LP time {[}s{]}} & \multicolumn{1}{c}{LP obj.} & \multicolumn{1}{c}{\#Non-zero} & \multicolumn{1}{c}{LP time {[}s{]}} & \multicolumn{1}{c}{LP obj.} \\ \midrule
\rowcolor[HTML]{EFEFEF} 
\cellcolor[HTML]{EFEFEF} & 0.25 & \cellcolor[HTML]{EFEFEF}$493.66 \pm 103.89$ & $0.93 \pm 0.15$ & $6.57 \pm 0.31$ & \cellcolor[HTML]{EFEFEF}$\mathbf{273.38 \pm 59.89}$ & $\mathbf{0.91 \pm 0.11}$ & $\mathbf{6.36 \pm 0.31}$ & \cellcolor[HTML]{EFEFEF}$303.12 \pm 57.47$ & $1.44 \pm 0.14$ & $\mathbf{6.36 \pm 0.31}$ \\
\rowcolor[HTML]{EFEFEF} 
\cellcolor[HTML]{EFEFEF} & 0.50 & $447.60 \pm 100.60$ & \cellcolor[HTML]{EFEFEF}$0.93 \pm 0.15$ & $5.36 \pm 0.29$ & $\mathbf{252.56 \pm 56.00}$ & $\mathbf{0.90 \pm 0.12}$ & $5.20 \pm 0.28$ & $281.28 \pm 61.69$ & $1.47 \pm 0.17$ & $\mathbf{5.19 \pm 0.28}$ \\
\rowcolor[HTML]{EFEFEF} 
\multirow{-3}{*}{\cellcolor[HTML]{EFEFEF}1.0} & 0.75 & $355.12 \pm 80.27$ & $\mathbf{0.87 \pm 0.14}$ & $4.20 \pm 0.31$ & $\mathbf{214.54 \pm 47.74}$ & $0.88 \pm 0.09$ & $4.08 \pm 0.30$ & $231.18 \pm 44.66$ & $1.39 \pm 0.14$ & $\mathbf{4.06 \pm 0.30}$ \\
 & 0.25 & $493.90 \pm 103.92$ & $0.93 \pm 0.16$ & $13.13 \pm 0.62$ & $\mathbf{273.18 \pm 59.46}$ & $\mathbf{0.90 \pm 0.10}$ & $12.73 \pm 0.61$ & $303.34 \pm 57.82$ & $1.44 \pm 0.14$ & $\mathbf{12.72 \pm 0.61}$ \\
 & 0.50 & $448.62 \pm 101.99$ & $\mathbf{0.93 \pm 0.15}$ & $10.71 \pm 0.57$ & $\mathbf{252.40 \pm 55.82}$ & $\mathbf{0.93 \pm 0.12}$ & $10.40 \pm 0.56$ & $281.18 \pm 61.44$ & $1.49 \pm 0.16$ & $\mathbf{10.38 \pm 0.56}$ \\
\multirow{-3}{*}{2.0} & 0.75 & $355.38 \pm 80.42$ & $\mathbf{0.87 \pm 0.11}$ & $8.40 \pm 0.62$ & $\mathbf{214.54 \pm 47.74}$ & $0.88 \pm 0.11$ & $8.15 \pm 0.60$ & $231.22 \pm 44.60$ & $1.41 \pm 0.14$ & $\mathbf{8.12 \pm 0.60}$ \\
\rowcolor[HTML]{EFEFEF} 
\cellcolor[HTML]{EFEFEF} & \cellcolor[HTML]{EFEFEF}0.25 & $493.58 \pm 104.61$ & $\mathbf{0.92 \pm 0.16}$ & $19.70 \pm 0.93$ & $\mathbf{274.38 \pm 60.98}$ & $0.94 \pm 0.15$ & $19.09 \pm 0.93$ & $302.84 \pm 58.59$ & $1.42 \pm 0.15$ & $\mathbf{19.08 \pm 0.92}$ \\
\rowcolor[HTML]{EFEFEF} 
\cellcolor[HTML]{EFEFEF} & 0.50 & $449.70 \pm 101.94$ & $0.94 \pm 0.17$ & $16.07 \pm 0.86$ & $\mathbf{251.78 \pm 55.85}$ & $\mathbf{0.93 \pm 0.14}$ & $15.60 \pm 0.84$ & $282.10 \pm 62.11$ & $1.48 \pm 0.16$ & $\mathbf{15.57 \pm 0.83}$ \\
\rowcolor[HTML]{EFEFEF} 
\multirow{-3}{*}{\cellcolor[HTML]{EFEFEF}3.0} & 0.75 & $356.94 \pm 82.48$ & $\mathbf{0.86 \pm 0.11}$ & $12.60 \pm 0.94$ & $\mathbf{214.38 \pm 47.18}$ & $0.92 \pm 0.10$ & $12.23 \pm 0.90$ & $231.34 \pm 44.76$ & $1.41 \pm 0.14$ & $\mathbf{12.18 \pm 0.89}$ \\ \bottomrule
\end{tabular}
\end{table}
\end{landscape}

\subsection{Performance Comparison of ResourceTune with Other Algorithms}

As already mentioned, the ResourceTune algorithm was compared with two existing algorithms, i.e., greedy algorithm and GA.
For this comparison, we generated $50$ instances for each pair of scenario parameters $\utilization \in \{1.0, 2.0, 3.0\}$ and $\proportion \in \{0.25, 0.50, 0.75\}$.
Then, we let each algorithm to generate 100 consecutive tuning plans for each problem instance, i.e., $|\plans| = 100$.

The resulting objective values are shown in Table~\ref{tab:objective}.
As can be seen, the objective value increased with $\utilization$ for all three algorithms.
This can be expected since higher $\utilization$ makes the instance more difficult to solve.
For almost all scenarios considered, ResourceTune's mean objective value is significantly better than those of the greedy algorithm and GA.
The only exception occurs when $\utilization = 1.0$ and $\proportion = 0.75$; in this case, the greedy algorithm outperforms ResourceTune.
However, the absolute difference between them is minimal and, in practice, almost non-existent.
In most scenarios, GA performed considerably worse than the other algorithms.
This was expected, given that the algorithm's runtime was limited to $2$ seconds.

\begin{table}[]
\centering
\caption{Objective values and win counts of ResourceTune, greedy algorithm, and GA for different values of scenario parameters $\utilization$ and $\proportion$.}
\label{tab:objective}
\begin{tabular}{@{}rrrrrrrr@{}}
\toprule
\multicolumn{1}{l}{} & \multicolumn{1}{l}{} & \multicolumn{6}{c}{Objective $\obj$} \\ \cmidrule(l){3-8} 
\multicolumn{1}{c}{} & \multicolumn{1}{c}{} & \multicolumn{2}{c}{ResourceTune} & \multicolumn{2}{c}{Greedy algorithm} & \multicolumn{2}{c}{GA} \\ \cmidrule(l){3-4} \cmidrule(l){5-6} \cmidrule(l){7-8} 
\multicolumn{1}{c}{$\utilization$} & \multicolumn{1}{c}{$\proportion$} & \multicolumn{1}{c}{Mean $\pm$ SD} & \multicolumn{1}{l}{\#Win} & \multicolumn{1}{c}{Mean $\pm$ SD} & \multicolumn{1}{l}{\#Win} & \multicolumn{1}{c}{Mean $\pm$ SD} & \multicolumn{1}{l}{\#Win} \\ \midrule
\rowcolor[HTML]{EFEFEF} 
\cellcolor[HTML]{EFEFEF} & 0.25 & $\mathbf{0.007 \pm 0.020}$ & \textbf{36} & $0.048 \pm 0.122$ & 14 & $0.626 \pm 0.252$ & 0 \\
\rowcolor[HTML]{EFEFEF} 
\cellcolor[HTML]{EFEFEF} & 0.50 & $\mathbf{0.004 \pm 0.003}$ & \cellcolor[HTML]{EFEFEF}\textbf{26} & $0.020 \pm 0.043$ & 24 & $0.322 \pm 0.149$ & 0 \\
\rowcolor[HTML]{EFEFEF} 
\multirow{-3}{*}{\cellcolor[HTML]{EFEFEF}1.0} & 0.75 & $0.006 \pm 0.003$ & \cellcolor[HTML]{EFEFEF}7 & $\mathbf{0.005 \pm 0.016}$ & \textbf{43} & $0.069 \pm 0.052$ & 0 \\
 & 0.25 & $\mathbf{1.492 \pm 0.428}$ & \textbf{45} & $1.736 \pm 0.421$ & 5 & $2.435 \pm 0.624$ & 0 \\
 & 0.50 & $\mathbf{0.844 \pm 0.279}$ & \textbf{49} & $1.239 \pm 0.294$ & 1 & $1.698 \pm 0.419$ & 0 \\
\multirow{-3}{*}{2.0} & 0.75 & $\mathbf{0.237 \pm 0.169}$ & \textbf{50} & $0.596 \pm 0.210$ & 0 & $0.963 \pm 0.237$ & 0 \\
\rowcolor[HTML]{EFEFEF} 
\cellcolor[HTML]{EFEFEF} & 0.25 & $\mathbf{3.477 \pm 0.794}$ & \cellcolor[HTML]{EFEFEF}\textbf{47} & $3.850 \pm 0.703$ & 3 & $4.372 \pm 0.976$ & 0 \\
\rowcolor[HTML]{EFEFEF} 
\cellcolor[HTML]{EFEFEF} & 0.50 & $\mathbf{2.485 \pm 0.563}$ & \cellcolor[HTML]{EFEFEF}\textbf{49} & $3.125 \pm 0.497$ & 1 & $3.264 \pm 0.671$ & 0 \\
\rowcolor[HTML]{EFEFEF} 
\multirow{-3}{*}{\cellcolor[HTML]{EFEFEF}3.0} & 0.75 & $\mathbf{1.397 \pm 0.367}$ & \cellcolor[HTML]{EFEFEF}\textbf{50} & $2.277 \pm 0.319$ & 0 & $2.215 \pm 0.389$ & 0 \\ \bottomrule
\end{tabular}
\end{table}

Large standard deviations must be addressed to eliminate doubts about the ResourceTune algorithm's performance and the fairness of the comparison.
We identified differences in complexity among individual instances as the source of these deviations.
To validate the results, we counted how many times each algorithm performed best in a given scenario.
These counts are shown in Table~\ref{tab:objective} in columns labeled \#win.
As can be seen, ResourceTune won in $\approx96.6\%$ of instances when $\utilization$ was $2.0$ or $3.0$.
The win counts are not as straightforward for instances when $\utilization$ was equal to $1.0$ because the greedy algorithm scored a non-negligible number of wins.
Nevertheless, as previously mentioned, in these instances, both ResourceTune and the greedy algorithm produce nearly optimal results, rendering the difference between them insignificant.

To further support our claims, we computed normalized objective values, denoted by $\obj^*$, for each algorithm and instance. 
These values were calculated by dividing the objective values by the smallest value produced by any algorithm on a given instance.
The first, second, and third quartiles of $\obj^*$ are shown in Table~\ref{tab:normalized}.
ResourceTune was better than any of the two compared algorithms in all instances except when $\utilization = 1.0$ and $\proportion = 0.75$, as previously discussed.
In the most extreme case, when $\utilization = 2.0$ and $\proportion = 0.75$, ResouceTune outperformed other algorithms in half of the instances more than $2.7$ times.
Even when the performance gap was the smallest, when $\utilization = 3.0$ and $\proportion = 0.25$, the ResourceTune algorithm was better in half of the instances by $10.0\%$ and $26.6\%$ than greedy algorithm and GA, respectively.

\begin{table}[]
\centering
\caption{Quartiles of normalized objective values of ResourceTune, greedy algorithm, and GA for different values of scenario parameters $\utilization$ and~$\proportion$.}
\label{tab:normalized}
\begin{tabular}{@{}rrrrrrrrrrr@{}}
\toprule
\multicolumn{1}{l}{} & \multicolumn{1}{l}{} & \multicolumn{9}{c}{Normalized objective $\obj^*$} \\ \cmidrule(l){3-11}
\multicolumn{1}{l}{} & \multicolumn{1}{l}{} & \multicolumn{3}{c}{ResourceTune} & \multicolumn{3}{c}{Greedy algorithm} & \multicolumn{3}{c}{GA} \\ \cmidrule(l){3-5} \cmidrule(l){6-8} \cmidrule(l){9-11} 
\multicolumn{1}{c}{$\utilization$} & \multicolumn{1}{c}{$\proportion$} & \multicolumn{1}{c}{Q1} & \multicolumn{1}{c}{Q2} & \multicolumn{1}{c}{Q3} & \multicolumn{1}{c}{Q1} & \multicolumn{1}{c}{Q2} & \multicolumn{1}{c}{Q3} & \multicolumn{1}{c}{Q1} & \multicolumn{1}{c}{Q2} & \multicolumn{1}{c}{Q3} \\ \midrule
\rowcolor[HTML]{EFEFEF} 
\cellcolor[HTML]{EFEFEF} & 0.25 & \textbf{1.000} & \textbf{1.000} & \textbf{1.478} & \textbf{1.000} & 3.378 & 23.174 & 224.189 & 365.718 & 978.545 \\
\rowcolor[HTML]{EFEFEF} 
\cellcolor[HTML]{EFEFEF} & 0.50 & \textbf{1.000} & \textbf{1.000} & \textbf{3.076} & \textbf{1.000} & 1.525 & 7.488 & 112.098 & 182.477 & 422.488 \\
\rowcolor[HTML]{EFEFEF} 
\multirow{-3}{*}{\cellcolor[HTML]{EFEFEF}1.0} & 0.75 & 1.053 & 3.172 & 15.836 & \textbf{1.000} & \textbf{1.000} & \textbf{1.044} & 20.604 & 35.187 & 147.284 \\
 & 0.25 & \textbf{1.000} & \textbf{1.000} & \textbf{1.000} & 1.079 & 1.171 & 1.288 & 1.559 & 1.639 & 1.743 \\
 & 0.50 & \textbf{1.000} & \textbf{1.000} & \textbf{1.000} & 1.391 & 1.506 & 1.618 & 1.855 & 2.044 & 2.223 \\
\multirow{-3}{*}{2.0} & 0.75 & \textbf{1.000} & \textbf{1.000} & \textbf{1.000} & 2.280 & 2.732 & 3.746 & 3.346 & 4.365 & 6.277 \\
\rowcolor[HTML]{EFEFEF} 
\cellcolor[HTML]{EFEFEF} & 0.25 & \textbf{1.000} & \textbf{1.000} & \textbf{1.000} & 1.057 & 1.100 & 1.166 & 1.230 & 1.266 & 1.300 \\
\rowcolor[HTML]{EFEFEF} 
\cellcolor[HTML]{EFEFEF} & 0.50 & \textbf{1.000} & \textbf{1.000} & \textbf{1.000} & 1.182 & 1.247 & 1.371 & 1.276 & 1.313 & 1.371 \\
\rowcolor[HTML]{EFEFEF} 
\multirow{-3}{*}{\cellcolor[HTML]{EFEFEF}3.0} & 0.75 & \textbf{1.000} & \textbf{1.000} & \textbf{1.000} & 1.503 & 1.667 & 1.859 & 1.520 & 1.588 & 1.718 \\ \bottomrule
\end{tabular}
\end{table}


\section{Conclusion}
\label{sec:conclusion}

This paper investigated the currently almost unexplored area of PSSRM.
We formulated the problem of constructing tuning plans for PSSs with two types of tasks, i.e., track and survey.
To solve this problem, we introduced the ResourceTune algorithm, which optimizes the observed frequencies to realize multiple tasks simultaneously.
This is done by combining the introduced left-right heuristic with LP, which produces optimized receiver configurations.
Additionally, we show that the proposed algorithm has low asymptotic computational complexity. 
This claim is further supported by experiments in which ResourceTune reliably constructed tuning plans in under 2 seconds.
After tuning the parameters, we experimentally demonstrated the superiority of the left-right heuristic over alternative approaches to configuration construction.
Finally, we compared the proposed algorithm with the greedy algorithm, which is inspired by the typical RRM algorithm, and GA as described by \citet{suja_scheduling}, which is currently the state-of-the-art algorithm for construction of tuning plans for PSS.
The comparison showed that ResourceTune either achieves near-optimal results, i.e., $\Theta$ close to $0.0$, or significantly outperforms both algorithms in almost all of the considered scenarios.
Future research could further improve the algorithm by incorporating track priorities and probabilities describing when the emitters will be active.
The algorithm's theoretical properties might also be studied, such as the optimality of the left-right heuristic or the computational complexity of similar interval-based problems.

\section*{CRediT authorship contribution statement}

\textbf{Jan Pikman:} Conceptualization, Formal analysis, Investigation, Methodology, Software, Visualization, Writing – original draft, Writing – review \& editing.
\textbf{Přemysl Šůcha:} Conceptualization, Formal analysis, Methodology, Supervision, Writing – review \& editing.
\textbf{Jerguš Suja:} Conceptualization, Resources, Writing – review \& editing.
\textbf{Pavel Kulmon:} Conceptualization, Resources, Writing – review \& editing.
\textbf{Zdeněk Hanzálek:} Funding acquisition, Project administration, Supervision, Writing – review \& editing.

\section*{Acknowledgements}
This work was supported by the European Union under the ROBOPROX project (reg. no. CZ.02.01.01/00/22\_008/0004590) and the Grant Agency of the Czech Technical University in Prague, grant No. SGS25/144/OHK3/3T/13.


\appendix

\section{Compared Algorithms}
\label{app:algorithms}

Since the PSSRM area is almost unexplored, both algorithms compared with ResourceTune had to be adapted to our problem.
Both algorithms have the same task preprocessing step, including the configuration construction process, which differs from the left-right heuristic. 
The configurations are constructed in two ways.
The first way involves dividing the frequency band $[10000, 16000]$ evenly into configurations of shape $(100, 100, 100)$ and weight equal to $4$.
Each configuration is then duplicated, and the copy's weight is set to $1$.
Finally, the surveys are split into sub-surveys according to the boundaries of these configurations.
The second way of configuration construction concerns only tracks and is almost identical to the one used in ResourceTune, except the configurations are centered on the emitter, i.e., they are not shifted to the left and right.

\subsection{Greedy Algorithm}
\label{subapp:greedy}

The greedy algorithm works by maintaining information about how often each track and sub-survey was observed.
To accomplish this, each task $\task \in (\tracks \cup \subsurveys)$ has time balance $\balance_\task \in \mathbb{R}$, which is initialized to $\goalObservation_\task$.
Every time the configuration that observes $\task$ is inserted into a tuning plan, its balance is updated
\begin{equation}
    \balance_\task \leftarrow \balance_\task - |\timesteps|^{-1}.
\end{equation}
After each tuning plan is constructed, the time balance of every task $\task \in (\tracks \cup \subsurveys)$ is updated (regardless of how many times it is observed by the newly constructed tuning plan)
\begin{equation}
    \balance_\task \leftarrow \balance_\task + \goalObservation_\task.
\end{equation}
Consequently, a positive balance indicates that a task has not been observed enough, while a negative balance indicates that the task has been observed sufficiently.
This balancing approach was taken from a typical RRM algorithm called the time balancing algorithm \citep{stafford_mesar, butler_time_balancing}.
Similar to ResourceTune, the greedy algorithm constructs tuning plans by repeatedly inserting the highest-priority configuration until the tuning plan is full.
The priority $\priority_\configuration$ of configuration $\configuration$ is determined by the balances of the tasks it measures:
\begin{equation}
    \priority_\configuration = \sum_{\task \in (\tracks \cup \subsurveys): \, \configuration \text{ observes } \task} \max\{0, \balance_\task\}.
\end{equation}

\subsection{Genetic Algorithm}
\label{subapp:ga}

The GA is implemented in the same way as described by \citet{suja_scheduling}.
In other words, the frequency bands observed by each receiver at each time step are represented by a single variable, and the configurations generated during preprocessing are the possible values of that variable.
In addition, it must be noted that the initial population partially consisted of individuals that were constructed by another GA that only considered configurations whose weight is $4$.
The objective function of GAs was $\obj$, as defined in Equation~\eqref{eq:obj}. 
It considered all of the previously constructed tuning plans.
It is important to note that this objective function was different from the one used in \citet{suja_scheduling}, which could affect its performance.
However, the concept of tuning plan construction by selecting frequency bands remained the same.
Finally, the runtime of both GAs was limited to $1$ second, resulting in an overall runtime of $2$ seconds.

\bibliographystyle{elsarticle-harv} 
\bibliography{sources.bib}

@phdthesis{sherwani,
  title={Resource management in active-passive multifunction radar networks},
  author={Sherwani, Hashir},
  year={2018},
  school={UCL (University College London)}
}

@INPROCEEDINGS{shaghaghi_b&b,
  author={Shaghaghi, Mahdi and Adve, Raviraj S.},
  booktitle={2017 IEEE Radar Conference (RadarConf)}, 
  title={Task selection and scheduling in multifunction multichannel radars}, 
  year={2017},
  volume={},
  number={},
  pages={0969-0974},
  doi={10.1109/RADAR.2017.7944344}
}

@INPROCEEDINGS{shaghaghi_value_network,
  author={Shaghaghi, Mahdi and Adve, Raviraj S.},
  booktitle={2018 IEEE Radar Conference (RadarConf18)}, 
  title={Machine learning based cognitive radar resource management}, 
  year={2018},
  volume={},
  number={},
  pages={1433-1438},
  doi={10.1109/RADAR.2018.8378775}
}

@inproceedings{shaghaghi_mcts,
  title={Resource management for multifunction multichannel cognitive radars},
  author={Shaghaghi, Mahdi and Adve, Raviraj S and Ding, Zhen},
  booktitle={2019 53rd Asilomar conference on signals, systems, and computers},
  pages={1550--1554},
  year={2019},
  organization={IEEE}
}

@article{miranda_knowledge_rm,
  title={Knowledge-based resource management for multifunction radar: a look at scheduling and task prioritization},
  author={Miranda, Sergio and Baker, Chris and Woodbridge, Karl and Griffiths, Hugh},
  journal={IEEE Signal Processing Magazine},
  volume={23},
  number={1},
  pages={66--76},
  year={2006},
  publisher={IEEE}
}

@article{hashmi_ai_survey,
    author = {Hashmi, Umair Sajid and Akbar, Sunila and Adve, Raviraj and Moo, Peter W. and Ding, Jack},
    title = {Artificial intelligence meets radar resource management: A comprehensive background and literature review},
    journal = {IET Radar, Sonar \& Navigation},
    volume = {17},
    number = {2},
    pages = {153-178},
    doi = {https://doi.org/10.1049/rsn2.12337},
    url = {https://ietresearch.onlinelibrary.wiley.com/doi/abs/10.1049/rsn2.12337},
    eprint = {https://ietresearch.onlinelibrary.wiley.com/doi/pdf/10.1049/rsn2.12337},
    year = {2023}
}

@inproceedings{ding_survey_rrm,
  title={A survey of radar resource management algorithms},
  author={Ding, Zhen},
  booktitle={2008 Canadian Conference on Electrical and Computer Engineering},
  pages={001559--001564},
  year={2008},
  organization={IEEE}
}

@inproceedings{qu_est_mc,
  title={A radar task scheduling method using random shifted start time with the {EST} algorithm},
  author={Qu, Zhen and Ding, Zhen and Moo, Peter},
  booktitle={2019 IEEE Radar Conference (RadarConf)},
  pages={1--5},
  year={2019},
  organization={IEEE}
}

@ARTICLE{kulmon_suja_sensor,
  author={Kulmon, Pavel and Suja, Jerguš and Benko, Matej},
  journal={IEEE Transactions on Radar Systems}, 
  title={Scheduling of Multi-Function Sensor}, 
  year={2023},
  volume={1},
  number={},
  pages={729-739},
  doi={10.1109/TRS.2023.3335208}
}

@article{zhang_resource_parn,
  title={Resource saving based dwell time allocation and detection threshold optimization in an asynchronous distributed phased array radar network},
  author={Zhang, Haowei and Weijian, LIU and Xiao, YANG},
  journal={Chinese Journal of Aeronautics},
  volume={36},
  number={11},
  pages={311--327},
  year={2023},
  publisher={Elsevier}
}

@phdthesis{butler_time_balancing,
    author={Joe M. Butler},
    title={Tracking and control in multi-function radar},
    school={University College London},
    year={1998}
}

@article{orman_scheduling_coupled_jobs,
  title={Scheduling for a multifunction phased array radar system},
  author={Orman, AJ and Potts, Chris N and Shahani, AK and Moore, AR},
  journal={European Journal of operational research},
  volume={90},
  number={1},
  pages={13--25},
  year={1996},
  publisher={Elsevier}
}

@ARTICLE{shi_transmit,
  author={Shi, Chenguang and Wang, Yijie and Salous, Sana and Zhou, Jianjiang and Yan, Junkun},
  journal={IEEE Transactions on Aerospace and Electronic Systems}, 
  title={Joint Transmit Resource Management and Waveform Selection Strategy for Target Tracking in Distributed Phased Array Radar Network}, 
  year={2022},
  volume={58},
  number={4},
  pages={2762-2778},
  doi={10.1109/TAES.2021.3138869}
}

@ARTICLE{shi_multidomain,
  author={Shi, Chenguang and Tang, Zhicheng and Ding, Lintao and Yan, Junkun},
  journal={IEEE Transactions on Aerospace and Electronic Systems}, 
  title={Multidomain Resource Allocation for Asynchronous Target Tracking in Heterogeneous Multiple Radar Networks With Nonideal Detection}, 
  year={2024},
  volume={60},
  number={2},
  pages={2016-2033},
  doi={10.1109/TAES.2023.3347214}
}

@ARTICLE{charlish_auction,
  author={Charlish, Alexander and Woodbridge, Karl and Griffiths, Hugh},
  journal={IEEE Transactions on Aerospace and Electronic Systems}, 
  title={Phased array radar resource management using continuous double auction}, 
  year={2015},
  volume={51},
  number={3},
  pages={2212-2224},
  doi={10.1109/TAES.2015.130558}
}

@ARTICLE{irci_qram_solutions,
  author={Irci, Ayhan and Saranli, Afsar and Baykal, Buyurman},
  journal={IEEE Transactions on Aerospace and Electronic Systems}, 
  title={Study on {Q-RAM} and Feasible Directions Based Methods for Resource Management in Phased Array Radar Systems}, 
  year={2010},
  volume={46},
  number={4},
  pages={1848-1864},
  doi={10.1109/TAES.2010.5595599}
}

@ARTICLE{nsga-ii,
  author={Deb, K. and Pratap, A. and Agarwal, S. and Meyarivan, T.},
  journal={IEEE Transactions on Evolutionary Computation}, 
  title={A fast and elitist multiobjective genetic algorithm: {NSGA-II}}, 
  year={2002},
  volume={6},
  number={2},
  pages={182-197},
  doi={10.1109/4235.996017}
}

@INPROCEEDINGS{gaafar_rl_mcts,
  author={Gaafar, Mohamed and Shaghaghi, Mahdi and Adve, Raviraj S. and Ding, Zhen},
  booktitle={2019 53rd Asilomar Conference on Signals, Systems, and Computers}, 
  title={Reinforcement Learning for Cognitive Radar Task Scheduling}, 
  year={2019},
  volume={},
  number={},
  pages={1653-1657},
  doi={10.1109/IEEECONF44664.2019.9048892}
}

@INPROCEEDINGS{stafford_mesar,
  author={Stafford, W.K.},
  booktitle={IEE Colloquium on Real-Time Management of Adaptive Radar Systems}, 
  title={Real time control of a multifunction electronically scanned adaptive radar ({MESAR})}, 
  year={1990},
  volume={},
  number={},
  pages={7/1-7/5},
  doi={}
}

@article{charlish_rrm,
  title={Array radar resource management},
  author={Charlish, Alexander and Katsilieris, Fotios and others},
  journal={Novel radar techniques and applications: real aperture array radar, imaging radar, and passive and multistatic radar},
  volume={1},
  pages={135--171},
  year={2017},
  publisher={Institution of Engineering and Technology}
}

@article{moo_scheduling_two_slope,
  title={Scheduling for multifunction radar via two-slope benefit functions},
  author={Moo, Peter W},
  journal={IET radar, sonar \& navigation},
  volume={5},
  number={8},
  pages={884--894},
  year={2011},
  publisher={IET}
}

@phdthesis{ing_thesis,
  title={Efficient scheduling for radar resource management.},
  author={Ing, Keith},
  year={2019},
  school={University of Melbourne, Parkville, Victoria, Australia}
}

@INPROCEEDINGS{vaillaud_simple,
  author={Vaillaud, Hugo and Hanen, Claire and Hyon, Emmanuel and Enderli, Cyrille},
  booktitle={2023 26th International Conference on Information Fusion (FUSION)}, 
  title={Target search with a radar on an airborne platform}, 
  year={2023},
  volume={},
  number={},
  pages={1-8},
  doi={10.23919/FUSION52260.2023.10224197}
}

@INPROCEEDINGS{vaillaud_overlap,
  author={Vaillaud, Hugo and Hanen, Claire and Hyon, Emmanuel and Enderli, Cyrille},
  booktitle={2023 18th Conference on Computer Science and Intelligence Systems (FedCSIS)}, 
  title={Target Search with an Allocation of Search Effort to Overlapping Cones of Observation}, 
  year={2023},
  volume={},
  number={},
  pages={801-811},
  doi={10.15439/2023F7181}}

@article{brown_iteration,
  title={Optimal search for a moving target in discrete time and space},
  author={Brown, Scott Shorey},
  journal={Operations research},
  volume={28},
  number={6},
  pages={1275--1289},
  year={1980},
  publisher={INFORMS}
}

@book{stone_optimal,
  title={Optimal search for moving targets},
  author={Stone, Lawrence D and Royset, Johannes O and Washburn, Alan R and others},
  year={2016},
  publisher={Springer}
}

@INPROCEEDINGS{vaidya_lp,
  author={Vaidya, P.M.},
  booktitle={30th Annual Symposium on Foundations of Computer Science}, 
  title={Speeding-up linear programming using fast matrix multiplication}, 
  year={1989},
  volume={},
  number={},
  pages={332-337},
  doi={10.1109/SFCS.1989.63499}
}

@INPROCEEDINGS{suja_scalarization,
  author={Suja, Jerguš and Kulmon, Pavel},
  booktitle={2024 New Trends in Signal Processing (NTSP)}, 
  title={Scalarization of Multi-Function Sensor Scheduling Problem}, 
  year={2024},
  volume={},
  number={},
  pages={1-5},
  doi={10.23919/NTSP61680.2024.10726299}
}

@ARTICLE{suja_scheduling,
  author={Suja, Jerguš and Kulmon, Pavel and Benko, Matej},
  journal={IEEE Transactions on Aerospace and Electronic Systems}, 
  title={Scheduling of multi-function multistatic sensor}, 
  year={2025},
  volume={},
  number={},
  pages={1-15},
  doi={10.1109/TAES.2025.3572871}
}

@Article{jiang_passive_location,
    AUTHOR = {Jiang, Jianjun and Zhang, Jing and Zhang, Lijia and Ran, Xiaomin and Tang, Yanqun},
    TITLE = {Passive Location Resource Scheduling Based on an Improved Genetic Algorithm},
    JOURNAL = {Sensors},
    VOLUME = {18},
    YEAR = {2018},
    NUMBER = {7},
    ARTICLE-NUMBER = {2093},
    URL = {https://www.mdpi.com/1424-8220/18/7/2093},
    PubMedID = {29966286},
    ISSN = {1424-8220},
    DOI = {10.3390/s18072093}
}

@misc{cbc,
  doi = {10.5281/ZENODO.13347261},
  url = {https://zenodo.org/doi/10.5281/zenodo.13347261},
  author = {Forrest, John and Ralphs, Ted and Vigerske, Stefan and Santos, Haroldo Gambini and Forrest, John and Hafer, Lou and Kristjansson, Bjarni and jpfasano and EdwinStraver and Jan-Willem and Lubin, Miles and rlougee and {a-andre} and jpgoncal1 and Brito, Samuel and {h-i-gassmann} and Cristina and Saltzman, Matthew and tosttost and Pitrus, Bruno and Matsushima, Fumiaki and Vossler, Patrick  and {Ron @ SWGY}   and to-st},
  title = {coin-or/Cbc: Release releases/2.10.12},
  publisher = {Zenodo},
  year = {2024},
  copyright = {Creative Commons Attribution 4.0 International}
}

@ARTICLE{pymoo,
    author={J. {Blank} and K. {Deb}},
    journal={IEEE Access},
    title={pymoo: Multi-Objective Optimization in Python},
    year={2020},
    volume={8},
    number={},
    pages={89497-89509},
}

@misc{era,
    author={{ERA a.s.}},
    year={2023},
    title={ERA Military Solutions},
    url={https://www.era.aero/downloads/presskit/company-military-2023.pdf},
    urldate={2025-11-03}
}

\end{document}